\documentclass{aa}
\usepackage{graphicx}
\usepackage{natbib}
\bibpunct{(}{)}{;}{a}{}{,} % to follow the A&A style
\usepackage{amsmath}	% Advanced maths commands
\usepackage{amssymb}	% Extra maths symbols 
\begin{document}

\title{Deep LOFAR 150 MHz imaging of the Bo\"otes field: Unveiling the faint low-frequency sky}

\titlerunning{Deep LOFAR observations of the Bo\"otes field}
\authorrunning{E. Retana-Montenegro et al.}

\author{E. Retana-Montenegro
  \inst{1}
  \and
  H. J. A. R\"ottgering
  \inst{1}
  \and
  T. W. Shimwell
  \inst{1}
  \and
  R. J. van Weeren
  \inst{1}
  \and
  I. Prandoni
  \inst{2}
  \and
  G. Brunetti
  \inst{2}
  \and
  P.N. Best
  \inst{3} 
  \and
  M. Br\"uggen
  \inst{4} 
  }

\institute{Leiden Observatory, Leiden University, P.O. Box 9513, 2300 RA, Leiden, The Netherlands
  \and
  INAF, Istituto di Radioastronomia, via P. Gobetti 101, 40129, Bologna, Italy
  \and
  SUPA, Institute for Astronomy, Royal Observatory, Blackford Hill, Edinburgh, EH9 3HJ, UK
  \and
  Hamburger Sternwarte, University of Hamburg, Gojenbergsweg 112, 21029 Hamburg, Germany
  }

\offprints{E. Retana-Montenegro, \email{edwinretana@gmail.com}}

% These dates will be filled out by the publisher
%\date{Accepted A\&A.} % slugcomment
%\date{Submitted to \aap; do not circulate.}
\date{\today}

%\keywords{surveys - catalogs \textendash{} radio continuum: general \textendash{}
%techniques: image processing }

%\keywords{surveys - catalogs - radio continuum: general - techniques: image processing}

\abstract{We have conducted a deep survey (with a central rms of $55\mu\textrm{Jy}$)
with the LOw Frequency ARray (LOFAR) at 120-168 MHz of the Bo\"otes
field, with an angular resolution of $3.98^{''}\times6.45^{''}$,
and obtained a sample of 10091 radio sources ($5\sigma$ limit) over
an area of $20\:\textrm{deg}^{2}$. The astrometry and flux scale
accuracy of our source catalog is investigated. The resolution bias,
incompleteness and other systematic effects that could affect our
source counts are discussed and accounted for. The derived 150 MHz
source counts present a flattening below sub-mJy flux densities, that
is in agreement with previous results from high- and low- frequency
surveys. This flattening has been argued to be due to an increasing
contribution of star-forming galaxies and faint active galactic nuclei.
Additionally, we use our observations to evaluate the contribution
of cosmic variance to the scatter in source counts measurements. The
latter is achieved by dividing our Bo\"otes mosaic into 10 non-overlapping
circular sectors, each one with an approximate area of $2\:\textrm{deg}^{2}.$
The counts in each sector are computed in the same way as done for
the entire mosaic. By comparing the induced scatter with that of counts
obtained from depth observations scaled to 150MHz, we find that the
$1\sigma$ scatter due to cosmic variance is larger than the Poissonian
errors of the source counts, and it may explain the dispersion from
previously reported depth source counts at flux densities $S<1\,\textrm{mJy}$.
This work demonstrates the feasibility of achieving deep radio imaging
at low-frequencies with LOFAR.} 

\keywords{surveys - catalogs - radio continuum: general - techniques: image processing}

\maketitle

\section{Introduction} 
The most luminous radio sources are often associated with radio-loud
active galactic nuclei (AGN) powered by accretion onto supermassive
black holes (SMBHs), whose radio emission is generated by the conversion
of potential energy into electromagnetic energy released as synchrotron
radiation and manifesting itself as large-scale structures (radio
jets and lobes). The less luminous radio-selected objects are mostly
associated with accreting systems like radio-quiet AGNs or starburst
galaxies. The radio-emission in star-forming systems has two components:
a non-thermal synchrotronic component produced by cosmic rays originating
from supernova shockwaves, and a thermal free-free component arising
from the interstellar medium ionization by hot massive stars \citep{1992ARAA..30..575C}.
Star formation is also thought to be responsible at least for a fraction
of radio emission in radio-quiet AGNs. \citep{2011ApJ...740...20P,2012ApJ...758...23C}. 

In recent years, many studies have confirmed a flattening
in the (Euclidean normalized) radio counts below a few mJy \citep{2008ApJS..177...14S,2009ApJ...694..235P}
first detected more than three decades ago \citep{1985ApJ...289..494W,1986HiA.....7..367K}.
This flatening is due to an increasing contribution of faint radio
sources at sub-mJy flux densities. The precise fraction associated
with different objects is still under debate, with studies showing
a mixture of ellipticals, dwarf galaxies, high-z AGNs, and starburst
galaxies \citep{2011MNRAS.411.1547P,2017A&A...602A...2S}. The plethora
of objects found suggests a complex interplay between star-formation
(SF) and AGN activity in the universe. 

\noindent Additional efforts are important to understand the physical
processes that trigger the radio emission of the sub-mJy and microJy
sources. Currently, this is partly hampered because the required sensitivity
to detect fainter objects have been achieved in only a few small patches
of the sky \citep{2010ApJS..188..384S,2012ApJ...758...23C,2013ApJS..205...13M,2016MNRAS.462.2934V,2017A&A...602A...1S}.

\noindent %The VLA-Cosmic Evolution Survey (COSMOS) 3 GHz Large Project bridges the gap between past and future radio continuum surveys by covering an area as large as two square degrees down to a sensitivity reached to date only for single %pointing observations. This allows for individual detections of > 10, 000 radio sources, further building on the already extensive radio coverage 

\noindent % The deep low-frequency radio surveys that are being carried out with the Low Frequency Array (LOFAR, ) will allow us to investigate in more detail the faint radio population.
%For instance, the fraction of optical AGNs with counterparts at low-frequencies is relatively higher than employing higher frequency imaging . This occurs because these objects are brighter at low-frequencies due to the steepening in %their spectral indexes. Moreover, the selection of bright radio sources at low-frequencies has proven successful in finding rare radio-galaxies at moderate and high redshifts . Another advantage is that the radio-band is a tracer of %SF activity, that is unaffected by the dust absorption . 

%This makes radio surveys a powerful diagnostic tool to understand the formation and evolution of the faintest and most distant radio sources. 
%than their higher frequency counterparts, as the flux density of these object is relatively higher at low-frequencies due to the steeping in their spectral indexes.%For all the foreamentioned reasons, the low-frequency radio sky remais still a parameter space largely unexplored, where important advances on the study of faint radio-sources can be achieved. 

\noindent The majority of deep surveys \citep{2010ApJS..188..384S,2012MNRAS.427.1830W,2013ApJS..205...13M,2016MNRAS.462.2934V,2017A&A...602A...1S}
have been carried using radio telescopes operating at high-frequencies
($>1.0$GHz). This situation is rapidly changing as the number of
low-frequency radio surveys ($<1.0$GHz) has increased in the last
few years. Some survey examples include the VLA Low frequency Sky
Survey (VLSS; \citealt{2007AJ....134.1245C}), Murchison Widefield
Array (MWA) Galactic and Extragalactic All-sky MWA survey (GLEAM;
\citealt{2015PASA...32...25W}), and the LOFAR Two-metre Sky Survey
(LoTSS, \citealt{2017AA...598A.104S}). However, several challenges
such as strong radio interference and varying effects like ionospheric
phase errors across the instrument field of view (FOV) make producing
high-resolution, low-frequency radio maps a difficult task \citep{2004SPIE.5489..817N}.
The necessity to overcome these challenges and to fully exploit the
science offered by low-frequency telescopes has spurred an invigorated
interest by radio-astronomers in improving the low-frequency calibration
and imaging techniques (e.g. \citealt{2004SPIE.5489..180C,2009A&A...501.1185I,2011MNRAS.414.1656K,2011A&A...527A.107S,2016ApJS..223....2V,2017arXiv171202078T}). 

% the First Alternative Data Release (ADR1) of the TIFR GMRT Sky Survey (TGSS; ),

\begin{figure}[tp]
\centering{}\includegraphics[bb=0bp 0bp 527bp 414bp,clip,scale=0.5]{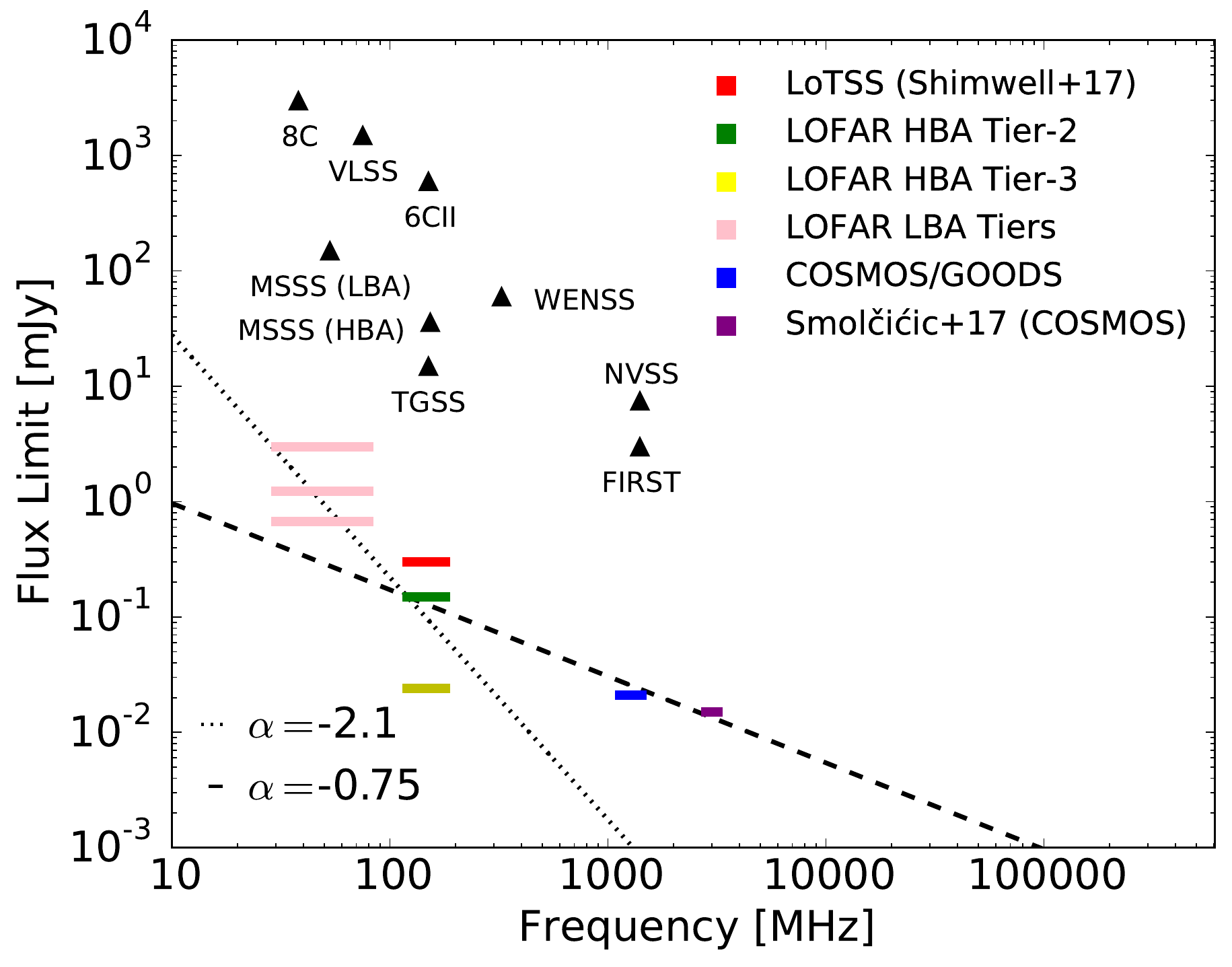}\centering\caption{\label{fig:lofar_tiers} Comparison between two radio sources with
the same flux, but different spectral indices. The black triangles
denote the \textbf{$5\sigma$} flux density limits for previous all-sky
shallow low- and high- frequency surveys \citep{1988MNRAS.234..919H,1995ApJ...450..559B,1998AJ....115.1693C,1997A&AS..124..259R,2007AJ....134.1245C,2015A&A...582A.123H,2017AA...598A..78I},
while color bars indicate the 3 different tiers for LOFAR surveys
using the LOFAR Low band antennas (LBA) and High band antennas (HBA),
and the deepest high-frequency surveys currently published \citep{2010ApJS..188..384S,2013ApJS..205...13M,2017A&A...602A...1S}.
Sources steeper than $\alpha=-2.1$ will be detected at higher signifcance
in the Tier2/Tier-3 surveys than in deep high-frequency surveys, while
sources flatter than $\alpha=-0.75$ at detected at both low and high
frequencies. }
\end{figure}

\noindent The LOFAR Surveys Key Science Project (SKSP) is embarking
on a survey with three tiers of observations: the LoTSS survey at
Tier-1 level covers the largest area at the lowest sensitivity ($\gtrsim100\,\mu$Jy)
covering the whole $2\pi$ steradians of the northern sky. Deeper
Tier-2 and Tier-3 programs aim to cover smaller fields with extensive
multi-wavelength data up to a depth of tens and a few microJy, respectively
(see \citealt{2011JApA...32..557R}). Together these surveys will
open the low-frequency electromagnetic spectrum for exploration, allowing
unprecedented studies of the faint radio population across cosmic
time and opening up new parameter space for searches for rare, unusual
objects such as high-z quasars \citep{10.3389/fspas.2018.00005} in
a systematic way (see Fig. \ref{fig:lofar_tiers}).

\noindent 

One of the regions for the Tier-2 and Tier-3 radio-continuum surveys
is the Bo\"otes field. This $9.2\:\textrm{deg}^{2}$ region is one
of the NOAO Deep Wide Field Survey (NDWFS, \citealt{1999ASPC..191..111J})
fields, and has a large wealth of multi-wavelength data available
including:\emph{ }X-rays (\emph{Chandra}; \citealt{2005ApJS..161....9K}),
optical (\emph{$U_{\textrm{spec}}$,}$B_{W}$,$R$,$I$\emph{,z,Y
}bands; \citealt{1999ASPC..191..111J,2007ApJS..169...21C,2013ApJ...774...28B}),
infrared ($J,H,K$\emph{ }bands\emph{, Spitzer}; \citealt{2003SPIE.4841..525A,2009ApJ...701..428A,2010AAS...21547001J})\emph{,
}and radio (60-1400 MHz; \citealt{2002AJ....123.1784D,2013AA...549A..55W,2014ApJ...793...82V,2016MNRAS.460.2385W}). 

\noindent In this work, we present deep 150 MHz LOFAR observations
of the Bo\"otes field obtained using the facet calibration technique
described by \citet{2016ApJS..223....2V}. The data reduction and
analysis for other deep fields using the kMS approach \citep{2014arXiv1410.8706T,2015MNRAS.449.2668S}
and \textsc{DDFacet} imager \citep{2017arXiv171202078T} will be presented
in future papers \citep{Mandal2018,Sabater2018,Tasse2018}. This paper
is structured as follows. In Sections \ref{sec:Section2} and \ref{sec:Section5},
we describe the observations and data reduction, respectevely. We
present our image and source catalog in Section \ref{sec:Section6}.
We also discuss for the flux density scale, astrometry accuracy, and
completeness and reliability. The differential source counts are presented
and discussed in Section \ref{sec:Section7}. The contribution of
cosmic variance to the scatter in source counts measurements is also
discussed in Section \ref{sec:Section7}. Finally, we summarise our
conclusions in Section \ref{sec:Section8}. We assume the convection
$S_{\nu}\propto\nu^{-\alpha}$, where $\nu$ is the frequency, $\alpha$
is the spectral index, and $S_{\nu}$ is the flux density as function
of frequency. 

% Also, we include a comparison with previous works.

\section{Observations\label{sec:Section2}}

The Bo\"otes observations centered at 14h32m00s +34d30m00s (J2000
coordinates) were obtained with the LOFAR High band antenna (HBA).
We combine 7 datasets observed from March 2013 (Cycle 0) to October
2015 (Cycle 4), which correspond aproximately to a total observing
time of 55 hours. When the LOFAR stations operate in the ``HBA DUAL
INNER'' configuration at 150 MHz, LOFAR has a half-power beam width
(HPBW) of $\sim5$ degrees with an angular resolution of $\sim5^{''}$
(using only the central and remote stations located in The Netherlands).
3C196 is used as primary flux calibrator and was observed 10 minutes
prior to the target observation. The nearby radio-loud quasar 3C295
was selected as secondary flux calibrator, and was observed for 10
minutes after the target. The observations from cycles 0 and 2 consist
of 366 subbands covering the range 110-182 MHz. The subbands below
120 MHz and above 167 MHz generally present poor signal\textendash to-noise
($S/N$). Therefore, in the following cycles, to obtain a more efficient
use of the LOFAR bandwidth the frequency range was restricted to 120-167
MHz, resulting in only 243 subbands per observation. The total time
on target varies depending on the cycle. The two observations from
Cycle 0 are 5 and 10 hours long, whereas Bo\"otes was observed for
8 hours per observation in Cycles 2 and 4. The frequency and time
resolution for the observations varies for each cycle. Table \ref{tab:summary_bootes}
presents the details for each one of the observations used in our
analysis. Our observations include the dataset L240772 analyzed by
\citet{2016MNRAS.460.2385W}.

%The frequency resolution for the observations varies from 4 channels per subband to 16 channels per subband, while the time resolution for cycle 2 and 4 observations is 2 and 1 seconds, respectively. Cycle 0 data has a time %resolution of 5 seconds. Table  presents the details for each one of the observations used in our analysis. Our observations include the dataset L240772 analyzed by .

%\begin{singlespace}

\begin{table*}
\caption{Summary of the LOFAR Bo\"otes observations. \label{tab:summary_bootes}}

\centering{}%
\begin{tabular}{ccccccccc}
 &  &  &  &  &  &  &  & \tabularnewline
\hline 
{\tiny{}Obs. ID } & {\tiny{}Amp. calibrator } & {\tiny{}Observation date} & {\tiny{}Frequency range} & {\tiny{}Subbands (Sb) } & {\tiny{}Ch. per sb} & {\tiny{}Ch. width} & {\tiny{}Int. time} & {\tiny{}Total time }\tabularnewline
 & {\tiny{} } &  & {\tiny{}{[}MHz{]} } &  &  & {\tiny{}{[}MHz{]} } & {\tiny{}{[}Seconds{]} } & {\tiny{}{[}Hours{]} }\tabularnewline
\hline 
{\tiny{}L240772} & {\tiny{}3C196} & {\tiny{}2014\textendash 08-10} & {\tiny{}110-182} & {\tiny{}000-365} & {\tiny{}8} & {\tiny{}24.41} & {\tiny{}2} & {\tiny{}8 }\tabularnewline
{\tiny{}L243561} & {\tiny{}3C196} & {\tiny{}2014\textendash 09-15} & {\tiny{}110-182} & {\tiny{}000-365} & {\tiny{}8} & {\tiny{}24.41} & {\tiny{}2} & {\tiny{}8 }\tabularnewline
{\tiny{}L374583} & {\tiny{}3C196} & {\tiny{}2014\textendash 09-24} & {\tiny{}120-169} & {\tiny{}244-487} & {\tiny{}16} & {\tiny{}12.21} & {\tiny{}1} & {\tiny{}8 }\tabularnewline
{\tiny{}L400135} & {\tiny{}3C196} & {\tiny{}2015\textendash 10-10} & {\tiny{}120-169} & {\tiny{}244-487} & {\tiny{}16} & {\tiny{}12.21} & {\tiny{}1} & {\tiny{}8 }\tabularnewline
{\tiny{}L401825} & {\tiny{}3C196} & {\tiny{}2015\textendash 10-21} & {\tiny{}120-169} & {\tiny{}244-487} & {\tiny{}16} & {\tiny{}12.21} & {\tiny{}1} & {\tiny{}8 }\tabularnewline
{\tiny{}L133895} & {\tiny{}3C196} & {\tiny{}2013\textendash 05-13} & {\tiny{}110-182} & {\tiny{}000-365} & {\tiny{}4} & {\tiny{}48.83} & {\tiny{}5} & {\tiny{}5 }\tabularnewline
{\tiny{}L131784} & {\tiny{}3C196} & {\tiny{}2013\textendash 05-07} & {\tiny{}110-182} & {\tiny{}000-365} & {\tiny{}4} & {\tiny{}48.83} & {\tiny{}5} & {\tiny{}10 }\tabularnewline
\hline 
\end{tabular}
\end{table*}

%\end{singlespace}

%The  from star-forming galaxies has been studied by.
%The total ex- tragalactic flux has been computed by adding the fluxes of all the sources which are not stars. In our diagnostic, stars fall at the bottom of the colour diagram and can be easily rverejected. Figure 4 reports the %fraction of AGN flux from the different fields in several flux bins. Poissonian error bars ar e reported. Due to the low number of sources, errors are still rather large but all the measurements are consistent and give an %estimate of the AGN contribution between 20% and 30%. The fraction is larger than the estimate done in the Lockman Hole using only X-ray identified AGN by Fadda et al. (2002) and is closer to more recent estimates based on %Spitzer 24

\section{Data reduction\label{sec:Section5}}

In this section, the data reduction steps of the LOFAR data processing
are briefly explained. These steps are divided into three stages:
the calibration into a non-directional and directional-dependent parts,
and the combination of the final calibrated datasets. We refer the
reader to the works of \citet{2016ApJS..223....2V} and \citet{2016MNRAS.460.2385W}
for a more detailed explanation of the calibration procedure. 

\subsection{Direction independent calibration}

First, we start by downloading the unaveraged data from the LOFAR
Long Term Archive (LTA)\footnote{http://lofar.target.rug.nl/}. We
follow the basic sequence of steps for the direction-independent (DI)
calibration: basic flagging and RFI removal employing \texttt{AOflagger}
\citep{2010MNRAS.405..155O,2012A&A...539A..95O}; flagging of the
contributing flux associated to bright off-axis sources referred as
the A-team (Cyg A, Cas A, Vir A, and Tau A); obtaining XX and YY gain
solution towards the primary flux calibrator using a 3C196 skymodel
provided by V.N. Pandey; determining the clock offsets between core
and remote stations using the primary flux calibrator phases solutions
as described by \citet{2016ApJS..223....2V}; measuring the XX and
YY phase offsets for the calibrator; transferring of amplitude, clock
values and phase offsets to the target field; averaging each subband
to a resolution of 4 seconds and 4 channels (no averaging is done
for cycle 0 data); initial phase calibration of the amplitude corrected
target field using a LOFAR skymodel of Bo\"otes. The final products
from the DI calibration are fiducial datasets consisting of 10 subbands
equivalent to 2 MHz bands. Each observation is composed of 23 or 21
bands depending on the number of bands flagged due to RFI. We limit
the frequency to the range 120-167 MHz to accomplish an uniform coverage
in the frequency domain. 

The DI calibrated bands are imaged at medium-resolution ($\sim40^{''}\times30^{''}$)
using \texttt{wsclean}\footnote{https://sourceforge.net/projects/wsclean/}
\citep{2014MNRAS.444..606O}. From these images, we construct a medium-resolution
skymodel that is subtracted from the visibility data. Later, these
data are imaged at low-resolution ($\sim110^{''}\times93^{''}$) to
obtain a low-resolution skymodel. This two-stage approach allows to
include extended emission that could have been missed in the medium-resolution
image. Both medium- and low- resolution skymodels are combined to
create the band skymodel. Finally, the band skymodel is used to subtract
the sources from the UV data to obtain DI residual visibilities. This
subtraction is temporarily, as these sources will be added later in
the directional self-calibration process. This stage of the data processing
is carried out using the \texttt{prefactor}\footnote{https://github.com/lofar-astron/prefactor/ }
tool. 

%Finally, all the sources included in the band skymodel are subtracted using the DI solutions from the UV data temporarily to obtain the DI residual visibilities, as a prior step to the directional self-calibration process. Theses %steps are applied for each observation using the pipeline processing tool for LOFAR HBA data prefactor. 

%cycle2:110.7422, 182.0312 chw=24414.06Hz
%cycle4:120.3125, 167.7734 chw= 12207.031250
%cycle0:ch_w=48828.125000Hz
%run02: pipelineid?133891?133895

\subsection{Direction dependent calibration}

\noindent 
\begin{figure}[tp]
\centering{}\includegraphics[clip,scale=0.31]{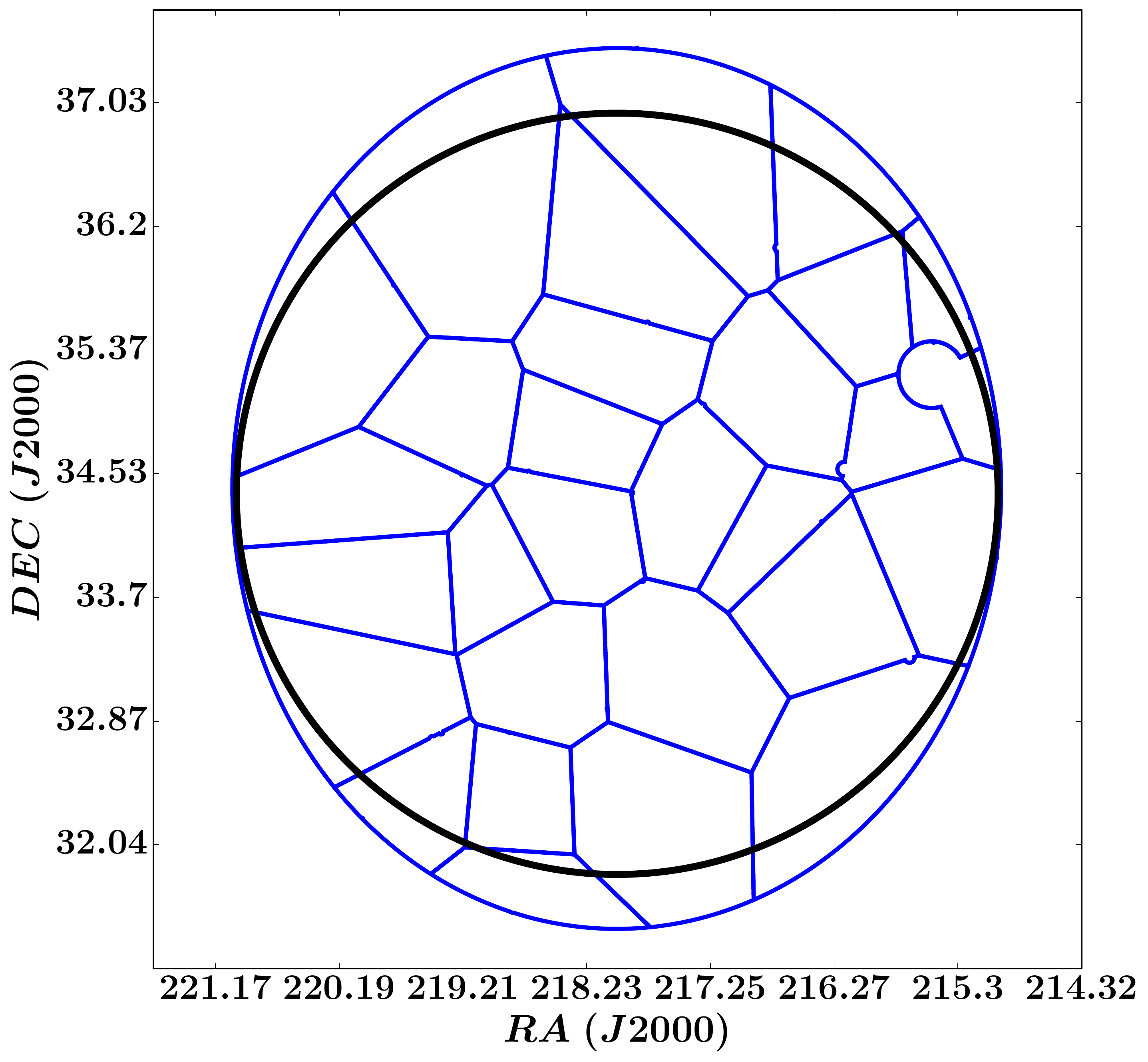}\centering\caption{\label{fig:facet_distribution} The spatial distribution of the facets
in the Bo\"otes field (blue solid lines). The large circle (solid
black line) indicates the radial cutoff of $2.5\:\textrm{degrees}$
used to apply the primary beam correction.}
\end{figure}

\noindent 
\begin{figure*}[t]
\begin{centering}
\includegraphics[clip,scale=0.9]{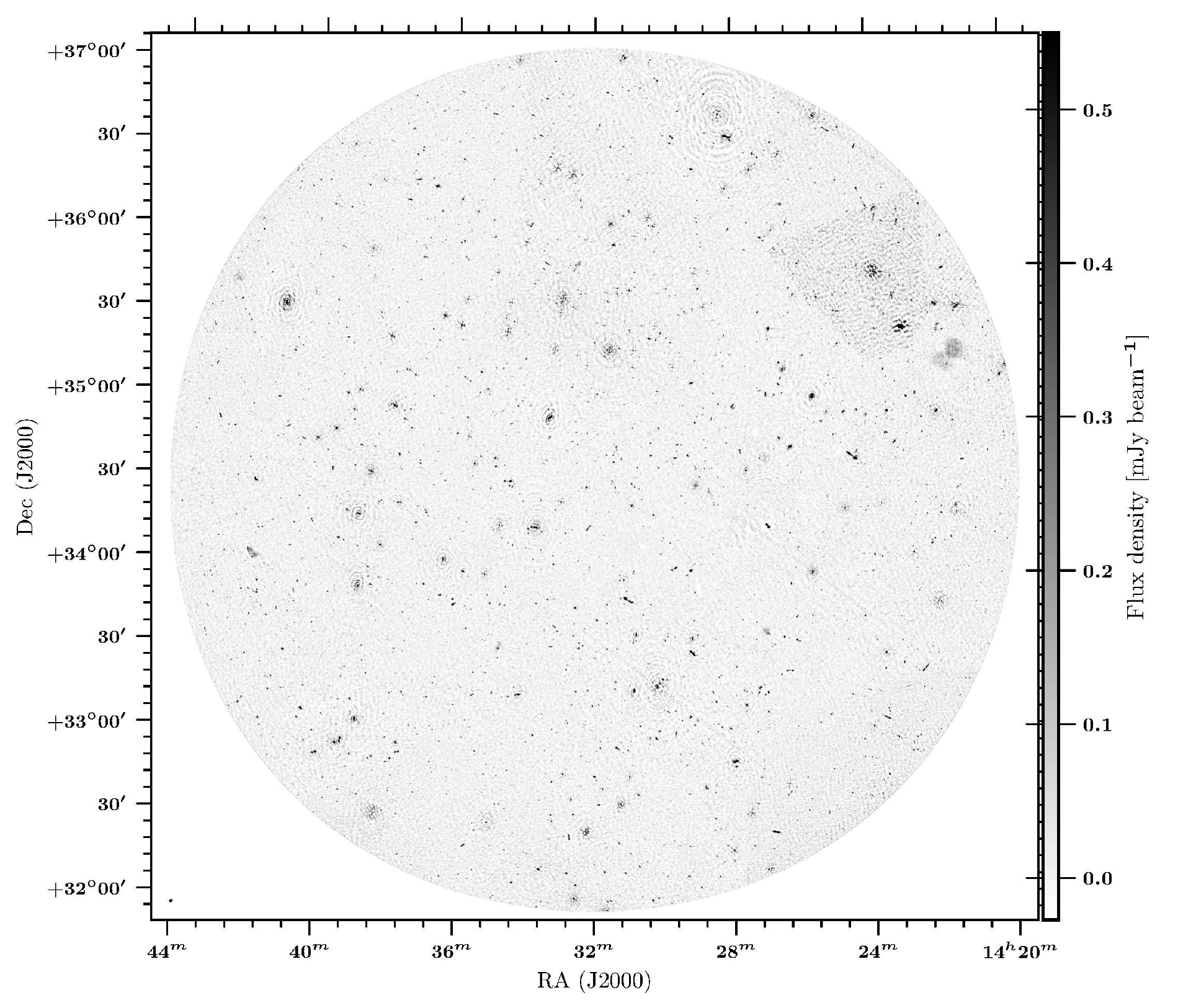}
\par\end{centering}
\centering{}\caption{\label{fig:full_mosaic_bootes} LOFAR 150 MHz mosaic of the Bo\"otes
field after beam correction. The size of the mosaic is approximately
$20\:{\textstyle \textrm{deg}^{2}}$. The synthesised beam size is
$5.5^{\prime\prime}\times7.4^{\prime\prime}$. The color scale varies
from $-0.5\sigma_{c}$ to $10\sigma_{c}$, where $\sigma_{c}=55\:\mu\textrm{Jy}/\textrm{beam}$
is the rms noise in the central region.}
\end{figure*}

\noindent 
\begin{figure*}[t]
\begin{centering}
\includegraphics[bb=0bp 0bp 644bp 454bp,clip,scale=0.8]{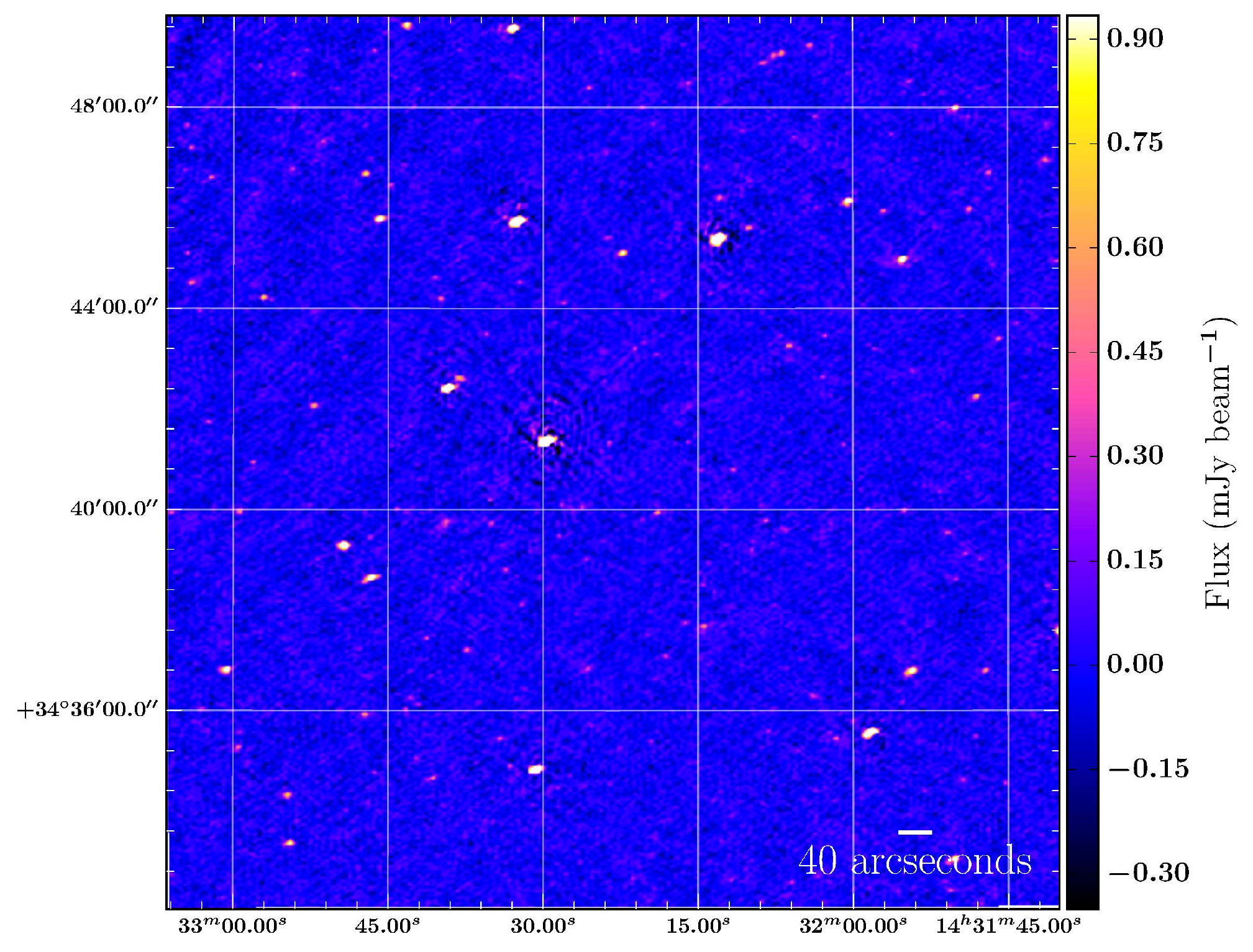}
\par\end{centering}
\centering{}\caption{\label{fig:central_bootes} Map showing the central $400^{\prime}\times400^{\prime}$
region of the mosaic center after primary beam correction. The synthesized
beam size is $5.5^{\prime\prime}\times7.4^{\prime\prime}$. The color
scale varies from $-6\sigma_{l}$ to $16\sigma_{l}$, where $\sigma_{l}=55\:\mu\textrm{Jy}/\textrm{beam}$
is the local rms noise.}
\end{figure*}

Direction-dependent (DD) effects such as the spatial and temporal
variability of the LOFAR station beam response, and the ionospheric
distortions must be considered to obtain high-fidelity low-frequency
radio images. It is well known that these effects are responsible
for artifacts and higher noise levels in low-frequency images (e.g.
\citealt{2013A&A...550A.136Y}). A simple approach to correct these
DD effects was originally proposed by \citet{1984AJ.....89.1076S}.
If the variation of the DD effects across the field of view (FOV)
is smooth, we can divide the FOV into a discrete number of regions
or ``facets''. Within each facet, there needs to be a bright source
or group of closely spaced bright sources, which is designated as
the facet calibrator. A self-calibration process can be performed
on each facet calibrator. This yields a set of DD calibration solutions
that are used to calibrate the whole facet. With the DD solutions
applied an image of the facet is made and a model for the sources
is created. Subsequently, this model is subtracted from the visibility
data, and the next brightest facet is dealt with \citep{2004SPIE.5489..817N}.
By executing these steps in an iterative way, it is possible to correct
the DD effects for all the facets in the FOV. Here, we adopt the DD
calibration technique described by \citet{2016ApJS..223....2V} to
process LOFAR HBA datasets. This procedure is now implemented in the
\texttt{factor}\footnote{https://github.com/lofar-astron/factor }
pipeline.

In our data processing, we use the same facet calibrator distribution
as \citet{2016MNRAS.460.2385W} with new boundary geometry (see Fig.
\ref{fig:facet_distribution}). The range of the flux density for
our facet calibrators is between 0.3 mJy and 2 Jy. To start the DD
process, the corresponding facet calibrator, which was subtracted
at the end of the DI calibration is added back to the UV data, and
all the bands are phase-rotated in the direction of the calibrator.
The self-calibration process comprises several cycles. In the first
and second cycles, we solve for the phase-offsets and the total ionospheric
electron content (TEC) terms (which introduces a frequency-dependent
ionospheric distortion on the phases offsets) only on timescales of
$\sim10$ seconds. For the the third and fourth cycles, we initially
solve only for phase+TEC. Finally, we obtain phase+amplitude solutions
on large timescales ($>5$ minutes for bright calibrators) to mainly
capture the relative slow variations in the beam. The last self-calibration
cycle can be iterated various times until convergence is achieved.
This last iteration step helps to decrease the number of artifacts
around bright facet calibrators. 

The imaging of the facet starts when the sources not selected as facet
calibrators are added back to the UV data and the DD solutions are
applied. The facet is imaged in two stages with \texttt{wsclean }\citep{2014MNRAS.444..606O}.
First, it is imaged at high resolution ($\sim5^{''}$) to include
all the compact sources in a high-resolution facet skymodel. Secondly,
the brightest sources from the high-resolution skymodel are subtracted,
and the facet is imaged at low-resolution $(\sim25^{''})$ to obtain
a skymodel that includes diffuse emission that can be missed during
the high-resolution imaging step. Both high and low resolution models
are combined into a new updated skymodel for the facet that is subtracted
from the full data. This process does not only improve the DI residual
visibilities by reducing the effective noise in the UV data as the
source subtraction is performed now using the DD solutions, but also
suppresses the effect of the presence of bright calibrators on the
subsequent subtraction of fainter facets. The facets are processed
in a serial sequence, which is ordered in descending order according
to the facet calibrator flux density. 

%. The facets are processed in a serial sequence, ordered by the descending value of the facet calibrator flux density. Each DI dataset is processed separately using the factor tool. 
%but also improves the residuals for the facets that will be self-calibrated later. 
%Each facet image provided an updated sky model that was then subtracted from the full-resolution data with the corresponding direction-dependent solutions, thereby improving the residual data to which the subsequent facets were %added. This process was repeated until all facets had been calibrated and im- aged. The order in which the facets were handled is determined by the severity of the calibration artefacts in the direction-independent images, which %roughly corresponds to the brightness of the cali- 

\subsection{Combined facet imaging}

The procedure to combine different observations is summarized in the
following steps:

%Shifting to a common phase center: The phase solutions for a particular facet are different for each of the observations. Thus, the facet phase center is going to present slight differences among the observations. To account for %this, we phase-shift all the facet data to a common phase center.
\begin{enumerate}
\item Shifting to a common phase center: For each facet, the astrometry
ultimately depends on the precision of the calibration model of the
facet calibrator. This implies that the astrometry can be shifted
between different regions due to the differences in precision between
the models of facet calibrators. This also explains the reason why
the astrometry for the same facet is usually slightly shifted, compared
to that of other observations. To account for the astrometry offsets
between different observations, we phase-shift all the data corresponding
to the same facet to a common phase center.
\item Normalizing imaging weights: The data from cycle 0 (4ch,5s) has been
further time averaged in comparison with the data from cycles 2 and
4 (4ch,4s). Thus, the imaging weights of cycle 0 data are multiplied
by a factor of 1.25 to account for the extra time averaging.
\item Facet imaging: The phase-shifted datasets from all the observations
corresponding to a facet are imaged together with \texttt{wsclean}.
We use a pixel size of $1.5^{''}$, and a robust parameter of $-0.7$
to obtain a more uniform weighting between short and remote baselines.
\item Mosaicing and primary-beam correction: The resulting facets from the
imaging step are mosaiced using \texttt{factor.} To apply the primary
beam correction, we use a beam model created by \texttt{wsclean}.
The correction is carried out by dividing the facet images by the
regridded \texttt{wsclean} beam model. We impose a radial cutoff where
the sensitivity of the phased array beam is 50 per cent of that at
the pointing center (i.e. a radius of $\sim2.5\:\textrm{deg}$).

%We notice that the visibilities from cycle 0 data present larger weights in comparison with posterior cycles. This results in a incorrect imaging, as the cycle 0 visibilites dominate over those of later cycles, which are ignored %by the imager. Therefore, for imaging purposes, we set constant weights of 1 to all visibilities in the datasets. 
%3Template mask: For each facet region, we create a template mask that delimits the cleaning to within the facet boundaries.%We follow the procedure described by  with a suitable modification for multi-epoch observations to apply the primary beam correction and create %the final mosaic. This includes applying an average primary-beam correction from all the observations obtained with AWimager . 
%. This includes convolving all the stacked facets to the same resolution to account for their different UV coverage
\end{enumerate}

\section{Images and sources catalog\label{sec:Section6}}

\noindent 
\begin{figure}
\begin{centering}
\includegraphics[clip,scale=0.45]{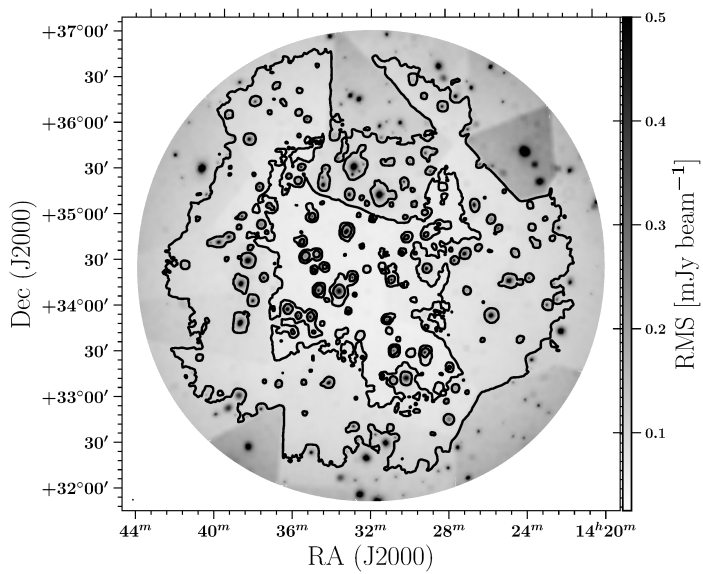}
\par\end{centering}
\centering{}\caption{\label{fig:full_rms_bootes} Noise map of the LOFAR 150 MHz mosaic
of the Bo\"otes field after primary beam correction. The color scale
varies from $0.5\sigma_{c}$ to $9\sigma_{c}$, where $\sigma_{c}=55\:\mu\textrm{Jy}/\textrm{beam}$
is the rms noise in the central region. Contours are plotted at $70\:\mu\textrm{Jy}/\textrm{beam}$
and $110\:\mu\textrm{Jy}/\textrm{beam}$.}
\end{figure}

\subsection{Final mosaic}

The final mosaic has an angular resolution of $3.98^{''}\times6.45^{''}$
with $\textrm{PA}=103^{\circ}$ and a central rms of $\sim55\:\mu\textrm{Jy}\textrm{/beam}$
. The entire mosaic and the central region of the Bo\"otes field
are shown in Fig. \ref{fig:full_mosaic_bootes} and Fig. \ref{fig:central_bootes},
respectively.

%Despite, a careful DD calibration some artifacts remain around bright sources caused by phase calibrationyu errors. These artifacts get accentuated in some cases for our deep stacked facet images, as the phase errors for the same %source from different observations get accumulated. However, these phase errors are confined to the surroundings of the bright objects. 

\subsection{Noise analysis and source extraction \label{sec:Section6-1}}

\noindent 
\begin{figure}[b]
\centering{}\includegraphics[clip,scale=0.55]{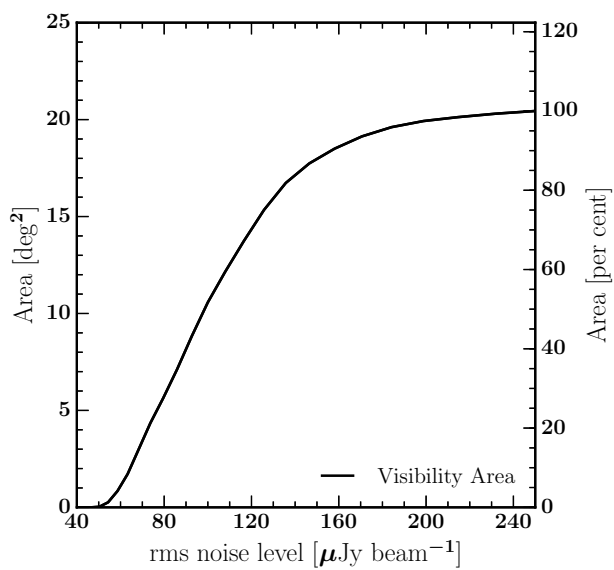}\centering\caption{\label{fig:visibility_area} Visibility area of the LOFAR image of
the Bo\"otes field. The full area covered is $20\:\textrm{deg}^{2}$.}
\end{figure}

\noindent We evaluate the spatial variation of the sensitivity of
our mosaic using a noise map created by \texttt{PyBDSF}\footnote{\texttt{https://github.com/lofar-astron/PyBDSF}}
\texttt{(}the Python Blob Dectection and Source Finder, formerly\texttt{
PyBDSM) }\citep{2015ascl.soft02007M}. The noise map of the Bo\"otes
mosaic is shown in Fig. \ref{fig:full_rms_bootes}. The noise threshold
varies from $\sim55\:\mu\textrm{Jy}/\textrm{beam}$ in the central
region to $\sim180\:\mu\textrm{Jy}/\textrm{beam}$ at the mosaic edges.
Around bright sources ($>500\:\textrm{mJy}/\textrm{beam}$), the image
noise can increase up to 5 times that of an unaffected region. This
is caused by residual phase errors still present after DD calibration.
The total area in which a source with a given flux can be detected,
or visibility area, of our mosaic is displayed in Fig \ref{fig:visibility_area}.
As expected, the visibility area increases rapidly between $\sim55\:\mu\textrm{Jy}/\textrm{beam}$
to $\sim250\:\mu\textrm{Jy}/\textrm{beam}$, with approximately 90
per cent of the mosaic area having a rms noise less than $160\:\mu\textrm{Jy}/\textrm{beam}$.
Two facets located near the mosaic edge have relatively higher noise
levels in comparison with adjacent facets. In these regions, the DD
calibration fails as their facet calibrators have low flux densities
($S_{150MHz}<1\textrm{mJy}$) resulting in amplitude and/or phase
solutions with low S/N ratios. The application of these poor solutions
to the data gives as result high-noise facets ($\sigma>120-150\:\mu\textrm{Jy}/\textrm{beam}$)
in the mosaic.

The software package \texttt{PyBDSF }was used to build an initial
source catalog within the chosen radial cutoff. The initial source
catalog consists of 10091 sources detected above a $5\sigma$ peak
flux density threshold. Of these 1978 are identified by \texttt{PyBDSF}
with the source structure code ``M'' (i.e. sources with multiple
components or complex structure), and the rest are classified as ``S''
(i.e. fitted by a single gaussian component). We inspected our mosaic
and found 170 multi-component sources that are misclassified into
different single sources by \texttt{PyBDSF} as their emission does
not overlap. This includes the 54 extended sources identified by \citet{2016MNRAS.460.2385W}.
The components for such sources are merged together by 1) assigning
the total flux from all the components as the total flux of the new
merged source, 2) assigning the peak flux of the brightest component
as the peak flux of the new merged source, and 3) computing the flux-weighted
mean position of the components and assigning it as the position of
the source. We list these merged sources as ``Flag\_merged'' in
the final source catalog. 

%Phase artifacts around high/intermediate flux density sources which were not fully removed during the DD calibration are still present in our stacked facets. Some of them could still %lead to spurious detections in our final catalog, despite, the source extraction is carried out with settings to properly mask phase errors as artifacts.

We visually inspected the surroundings of bright objects to identify
fake detections. A total of 119 objects are identified as artifacts
and flagged ``Flag\_artifact'' in our final catalog. These objects
are excluded from our source counts calculations (see Section \ref{sec:Section8}).

% Additionally, we visually inspect weak sources in the vicinity of bright to look for potential artifacts. In total, we identify 45 sources as 'Flag artifact' in the final catalog.

\subsection{Astrometry \label{sec:astrometry_correction}}

%Within the overlapping region between our maps and the FIRST survey,
%We exclude FIRST sources with the side lobe flags

%0.659933447521 0.626683706171 

\begin{figure*}[tp]
\begin{centering}
\includegraphics[clip,scale=0.45]{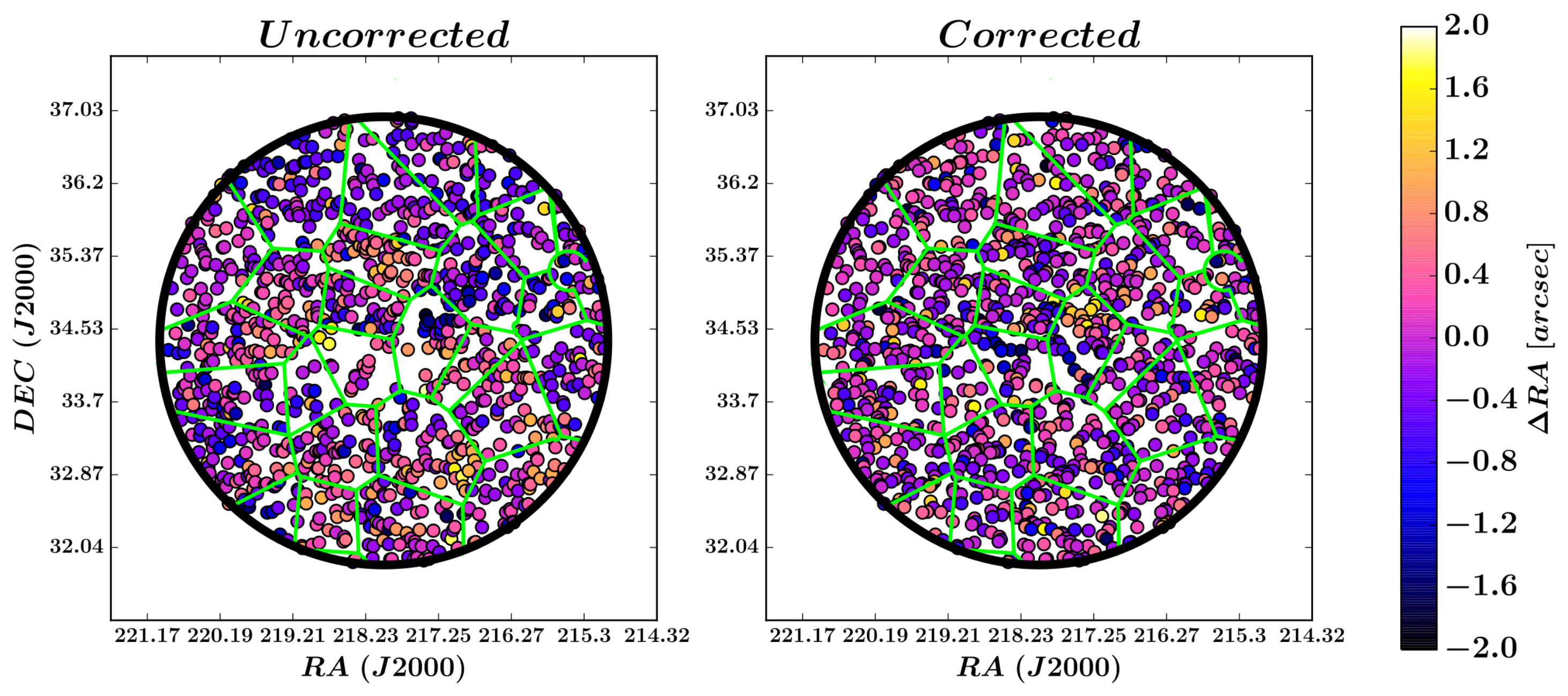}
\par\end{centering}
\begin{centering}
\includegraphics[clip,scale=0.45]{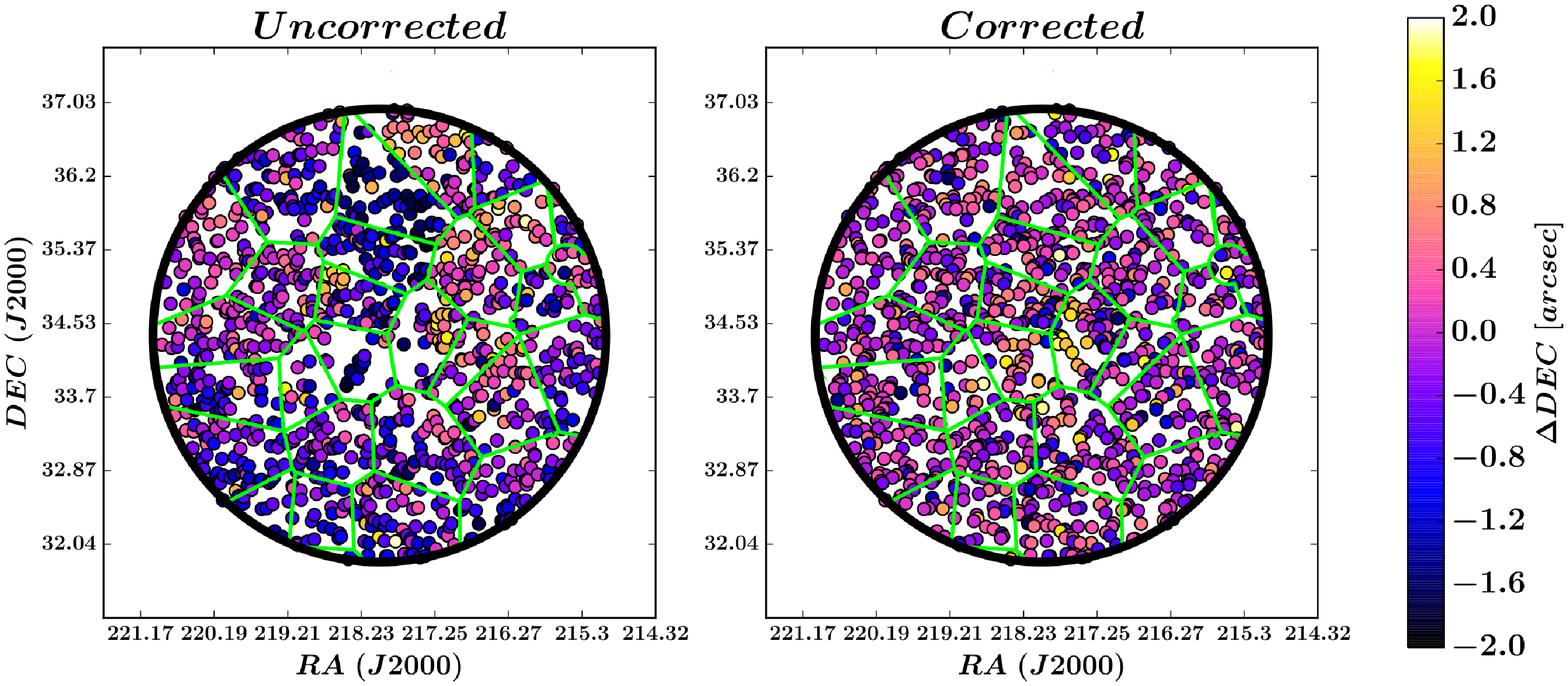}
\par\end{centering}
\centering{}\centering\caption{\label{fig:bootes_astrometry_mosaic} The spatial distribution of
positional offsets uncorrected (left) and corrected (right) between
high S/N and compact LOFAR sources and their FIRST counterparts in
the right ascention (top) and declination (bottom) directions. The
colorbar denotes the offsets for each object. We find 989 LOFAR/FIRST
sources (left panels) using the uncorrected positions; when the astrometry
corrections are applied a total of 1048 LOFAR/FIRST sources are found
(right panels). The black circle indicates the radial cutoff used
to apply the primary beam correction, while the green lines show the
facet distribution in our Bo\"otes mosaic.}
\end{figure*}

\begin{figure}[tp]
\centering{}\includegraphics[bb=0bp 0bp 579bp 579bp,clip,scale=0.45]{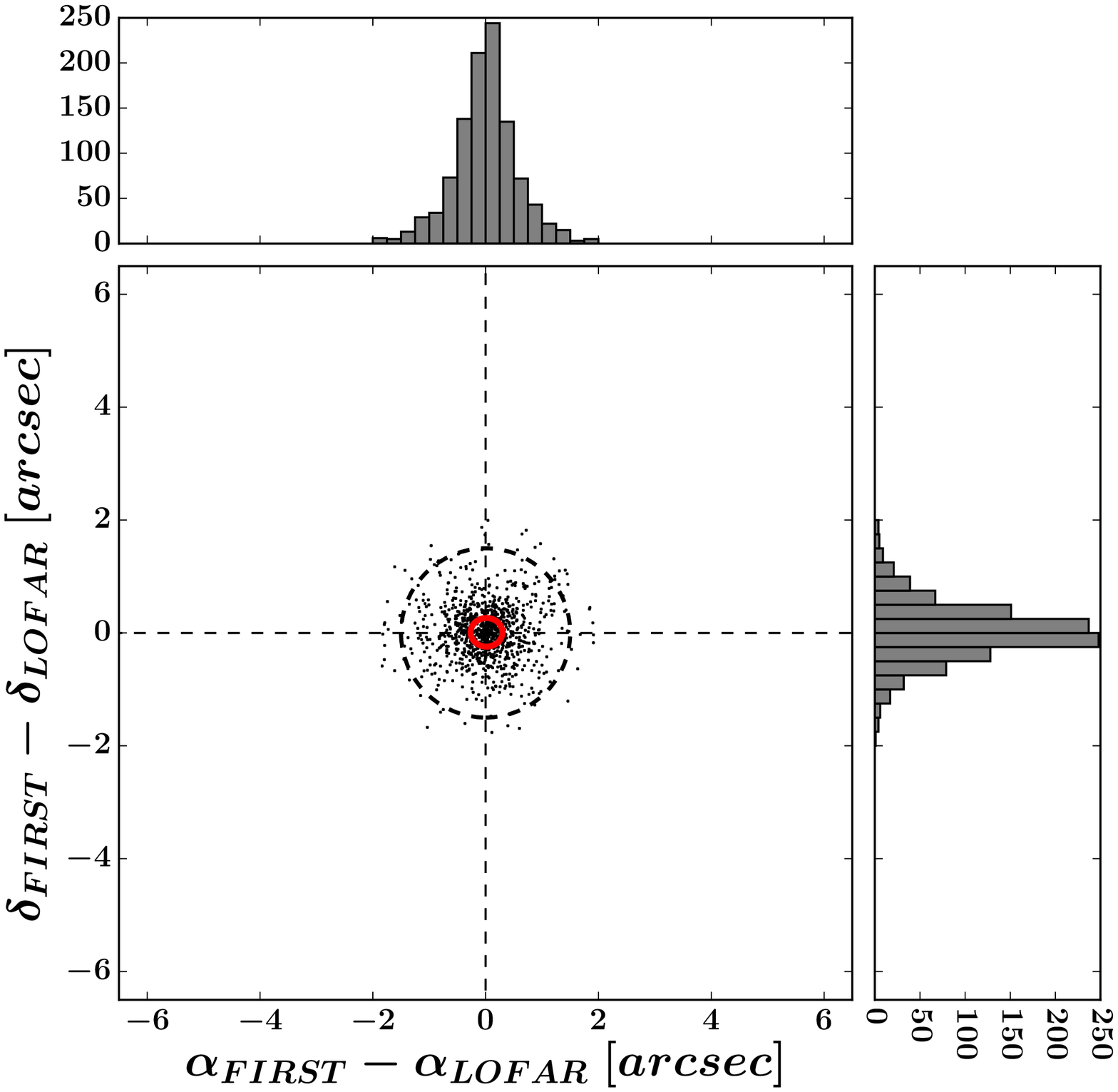}\centering\caption{\label{fig:bootes_astrometry} The corrected positional offsets between
high S/N and compact LOFAR sources and their FIRST counterparts (see
text for more details). The dashed lines denotes a circle with radius
$r=1.5^{\prime\prime},$ which is the image pixel scale. The ellipse
(red solid line) centered on the right ascension and declination mean
offsets indicates the standard deviation for both directions. }
\end{figure}

\noindent To check the positional accuracy, the LOFAR data is cross-correlated
against the FIRST survey \citep{1995ApJ...450..559B}. We crossmatched
the two catalogs using a matching radius of $2^{''}$. In order to
minimize the possibility of mismatching, we consider only LOFAR sources
with the following criteria: i) a $S/N>10$ in both LOFAR and FIRST
maps (i.e. high S/N sources), and ii) an angular size less than $50^{''}$
to select only compact sources with reliable positions. We find that
the mean offsets in right ascension and declination for the cross-matched
989 LOFAR sources are $\left\langle \alpha\right\rangle =0.012\pm1\times10^{-4}\;\textrm{arcsec}$
and $\left\langle \delta\right\rangle =0.27\pm1\times10^{-4}\;\textrm{arcsec}$,
respectively. The standard deviations of the right ascension and declination
are $\sigma_{RA}=0.57\;\textrm{arcsec}$ and $\sigma_{DEC}=0.64\;\textrm{arcsec}$,
respectively. The examination of the offsets in the right ascension
and declination directions shows that these have an asymmetrical distribution
that differs between facets (see Fig. \ref{fig:bootes_astrometry_mosaic},
left panels). We correct the positional offsets in both directions
using the FIRST catalog for each facet independently. This is done
by fitting a 2D plane to the offsets between the LOFAR and FIRST positions.
The plane is $A_{0}\left(\alpha-\alpha_{0}\right)+B_{0}\left(\delta-\delta_{0}\right)+C_{0}=0$,
where $\alpha$ and $\delta$ are the right ascension and declination
of the LOFAR-FIRST sources, respectively, $\alpha_{0}$ and $\delta_{0}$
are the central right ascension and declination of the corresponding
facet, and the constants $A_{0}$, $B_{0},$ and $C_{0}$ have units
of arcseconds. This fitting provides the astrometry correction that
is applied to all sources withing the corresponding facet (see Fig.
\ref{fig:bootes_astrometry_mosaic}, right panels). We find a total
of selected 1048 LOFAR/FIRST sources after the corrections are applied.
The mean offsets for the corrected positions are $\left\langle \alpha\right\rangle =0.009\pm1\times10^{-4}\;\textrm{arcsec}$
and $\left\langle \delta\right\rangle =0.005\pm3\times10^{-4}\;\textrm{arcsec}$,
respectively. The standard deviations are $\sigma_{RA}=0.42\;\textrm{arcsec}$
and $\sigma_{DEC}=0.40\;\textrm{arcsec}$, respectively. Fig. \ref{fig:bootes_astrometry}
shows the corrected positional offsets. As these offsets are typically
smaller than the pixel scale in our mosaic, we do not apply any further
corrections for positional offsets in our catalog. 

%The constants have median values of  arcseconds.
%Although these values are less than the pixel scale in our mosaic, we add in quadrature a  contribution to the errors in right ascension and declination to account for the uncertainty in the positional offsets.

%Additionally, the variation in declination found by  is not present in our images.

\noindent 

\noindent 

\subsection{Bandwidth and time smearing \label{sec:smearing_correction}}

Two systematic effects that must be accounted for are bandwidth and
time smearing. This combined smearing effect reduces the peak flux
of a source, and simultaneously the source size is distorted or blurred
in such way that the total flux is conserved, but the peak flux is
reduced. The smearing effect depends on resolution, channel width,
integration time, and increases with the source distance from the
phase center. \citet{2016MNRAS.460.2385W} averaged their data to
a resolution of 2 channels and 8 seconds, which yields a peak flux
decrease of 15 per cent at $2.5$ degrees from the pointing center
according to the equations given by \citet{1999ASPC..180..371B}.
In this work, the reduction in peak flux is less severe as our averaging
factor is two times smaller in frequency and time. This results in
a reduction of roughly 8 per cent at the same distance. This holds
for all the datasets that were not observed in cycle 0. For cycle
0 observations, the resolution available is 4 channel and 5 seconds.
In this case, the peak flux underestimation is approximately 10 per
cent at 2.5 deg from the pointing center. Following \citet{1999ASPC..180..371B},
we apply a weighted smearing correction that takes into account the
frequency resolution and integration time of the data sets. The factor
for Cycle 0 observations is $15/55=0.27$ (i.e. the ratio between
the observing time obtained in Cycle 0 and the total observing time),
and for the other cycles the factor is $40/55$ (i.e. its reciprocal
$0.73$). The smearing correction factor ($\geq1.0$) depends on the
distance of the source from the pointing center.

%Following , we apply a smearing correction that assigns a different weight to the different data resolutions based on its time contribution to the total observing time. The weight for the correction factor of the first resolution %(4ch4s) is , while for the second resolution (4ch5s) the weight is . This correction is applied, on a per object basis, as a function of the distance from the pointing center.

%Since all the datasets are combined together in the UV plane
%We have also incorporated the correction for time and bandwidth smearing, which causes a maxi- mum underestimation of the peak flux of 0.93. 

\subsection{Flux density scale accuracy \label{sec:flux_correction}}

%Despite, our flux density calibration has been performed with an accurate 3C196 skymodel, there is still a possibility that our resulting flux scale may not be consistent with other low-frequency surveys. Several reasons may explain %this difference: ionospheric blurring, differences in the flux density between our calibrator model and those predicted by  flux scale, the uncertainties on the LOFAR HBA synthesized beam model and the transfer uncertainties on gain %normalizations due to calibrator-target elevation differences.

% which indicates that our flux scale is consistent with the  scale

\begin{figure}[tp]
\centering{}\includegraphics[clip,scale=0.52]{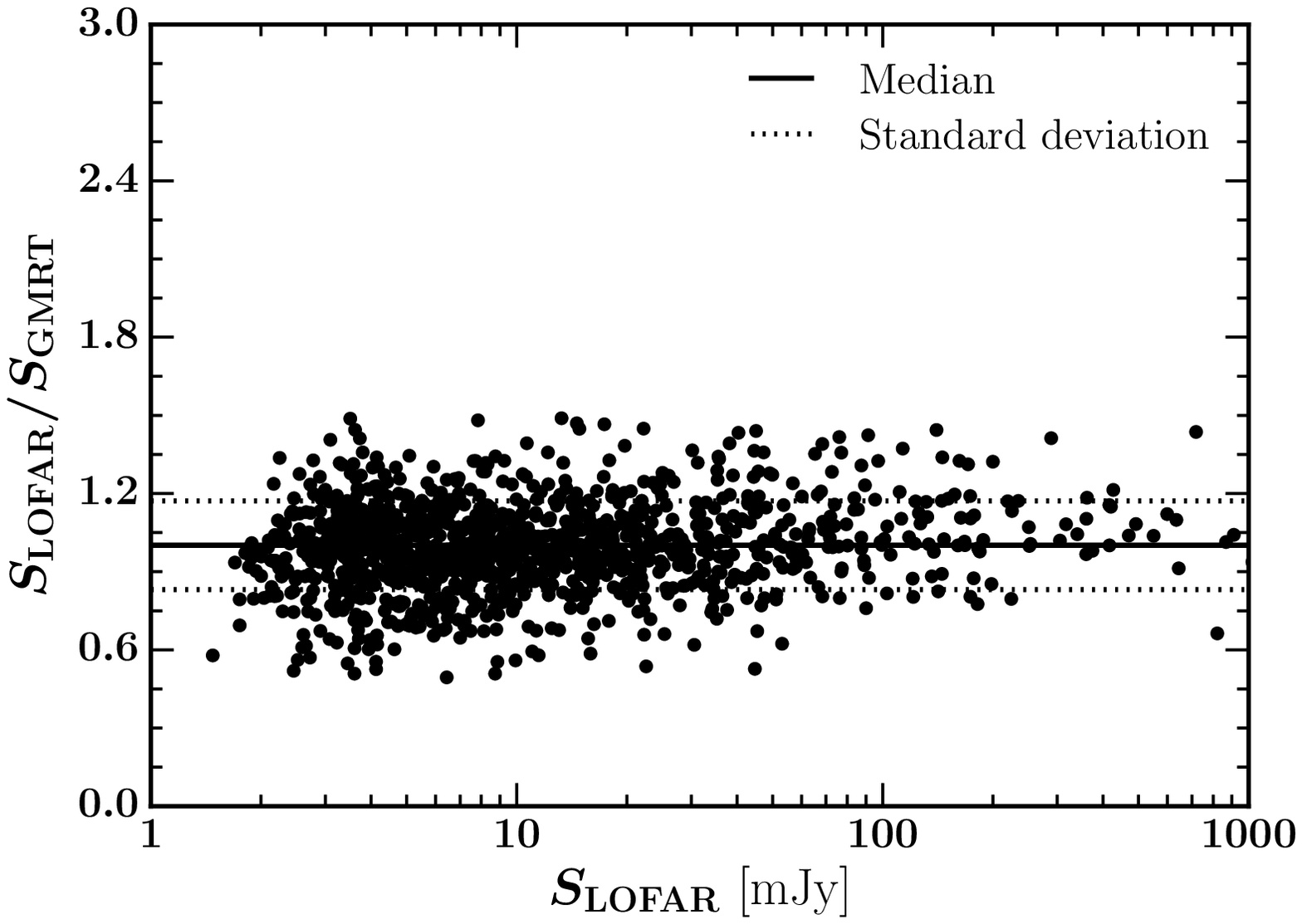}\centering\caption{\label{fig:flux_scale} Total flux ratio for LOFAR sources and their
GMRT counterparts. Only unresolved and isolated LOFAR sources with
$S/N>15$ are considered (see text for more details). The dashed lines
correspond to a standard deviation of $\sigma_{f_{R}}=0.12$, and
the median ratio of $1.00$ is indicated by a solid black line.}
\end{figure}

\noindent To verify the flux density scale for our Bo\"otes catalog
and check its consistency with the \citet{2012MNRAS.423L..30S} flux
scale, we compare our fluxes with the GMRT 150 MHz Bo\"otes catalog
by \citet{2013AA...549A..55W}. These authors obtained a mosaic with
rms levels of $2-5$mJy and an angular resolution of $25\;\textrm{arcsec}$.
First, a representative sample of sources is chosen using the following
criteria: i) a $S/N>15$ in both LOFAR and GMRT maps (i.e. high S/N
sources), ii) an angular size less than $50^{''}$, and iii) no neighbors
within a distance equal to the GMRT beam size or $25^{\prime\prime}$
(i.e. isolated sources). Secondly, we use a scaling factor of $1.078$
to put the GMRT fluxes on the \citet{2012MNRAS.423L..30S} scale,
according to the 3C196 calibration model \citep{2016MNRAS.460.2385W}.
The crossmatching yields a total of 1250 LOFAR/GMRT sources. We find
a mean flux ratio of $f_{R}=0.88$ with a standard deviation of $\sigma_{f_{R}}=0.15$,
which indicates a systematic offset in our flux scale in comparison
with the GMRT fluxes. Thus, we apply a correction factor of 12 per
cent to our LOFAR fluxes. After correcting the fluxes, we find a mean
flux ratio of $f_{R}=1.00$ with a standard deviation of $\sigma_{f_{R}}=0.12$
(see Fig. \ref{fig:flux_scale}). Considering uncertainties on the
flux scale such as: the accuracy of the fluxes on LOFAR images obtained
using skymodels based on the \citet{2012MNRAS.423L..30S} is approximately
of 10 per cent (e.g. \citealt{2016MNRAS.463.2997M,2017AA...598A.104S}),
the errors of the GMRT flux scale \citep{2016MNRAS.460.2385W}, and
the differences in elevation between the calibrator and target, we
conclude that a 15 per cent uncertainty in our flux scale is appropriate.
These global errors are added in quadrature to the flux uncertainties
reported by \texttt{PyBDSF} in our final catalog.

\noindent 

\subsection{Resolved sources \label{sec:resolved_sources}}

\noindent We estimate the maximum extension of a radio source using
the total flux $S_{T}$ to peak flux $S_{P}$ ratio:
\noindent \begin{center}
\begin{equation}
S_{T}/S_{P}=\theta_{\textrm{maj}}\theta_{\textrm{min}}/b_{\textrm{min}}b_{\textrm{maj}},\label{eq:St/Sp_vs_angle}
\end{equation}
\par\end{center}

\noindent where $\theta_{\textrm{min}}$ and $\theta_{\textrm{maj}}$
are the source FWHM axes, $b_{\textrm{min}}$ and $b_{\textrm{maj}}$
are the synthesized beam FWHM axes. The correlation between the peak
and total flux errors produces a flux ratio distribution with skewer
values at low S/N, while it has a tail due to extented sources that
extends to high ratios \citep{2000A&AS..146...41P}. If $S_{T}/S_{P}<1$
sources are affected by errors introduced by the noise in our mosaic,
we can derive a criterion for extension assuming that these errors
affect $S_{T}/S_{P}>1$ sources as well. The lower envelope (the curve
that contains 90 per cent of all sources with $S_{P}<S_{T}$) is fitted
in the $S_{P}/\sigma$ axis (where $\sigma$ is the local rms noise).
This curved is mirrored above the $S_{P}=S_{T}$ axis, and is described
by the equation:
\noindent \begin{center}
\begin{equation}
S_{T}/S_{P}=1.09+\left[\frac{2.7}{(S_{P}/\sigma)}\right].\label{eq:upper_envelope}
\end{equation}
\par\end{center}

\begin{figure}
\centering{}\centering\includegraphics[scale=0.55,clip=true]{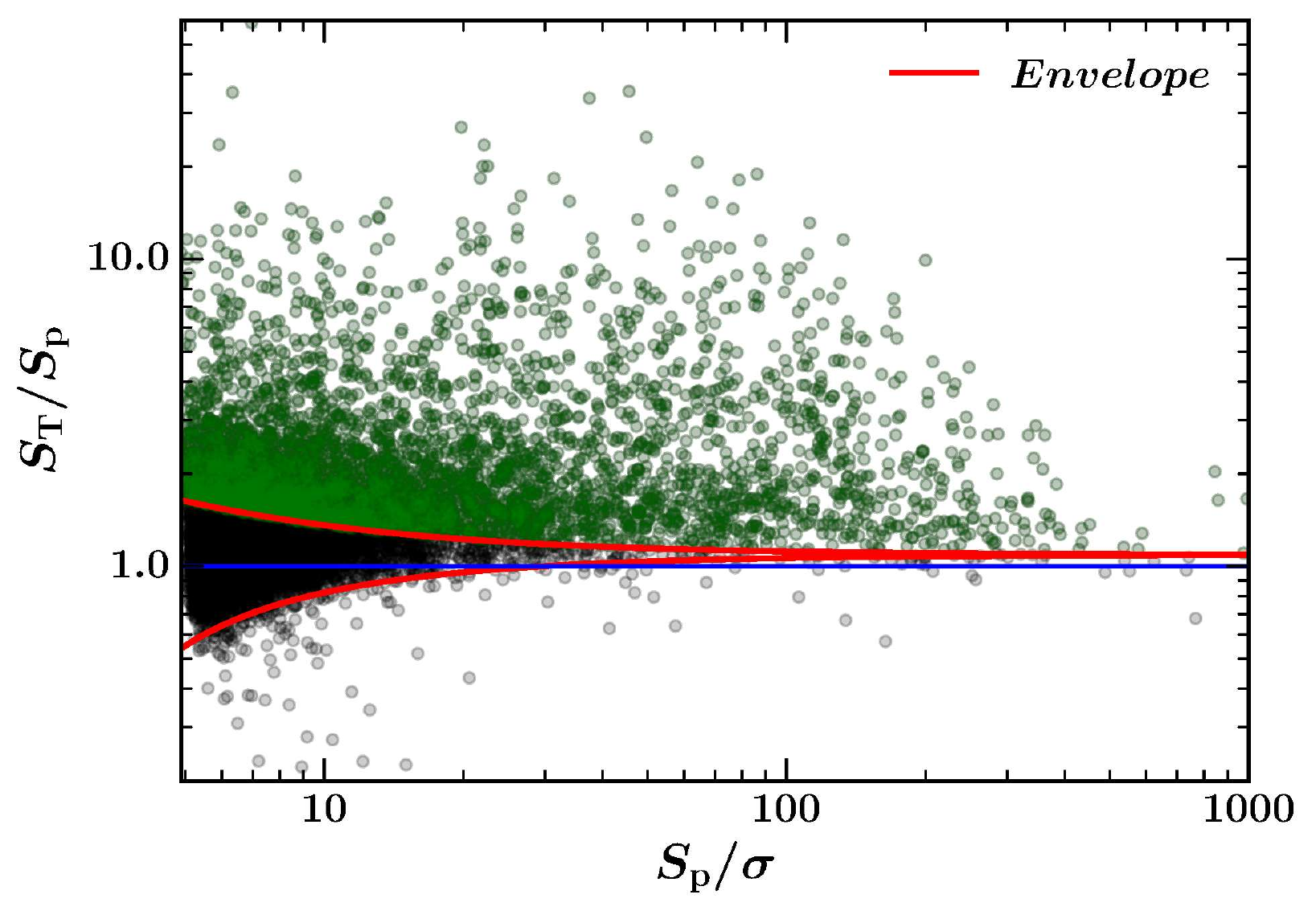}\caption{\label{fig:simulations_real} Ratio of the total flux density $S_{T}$
to peak flux density $S_{P}$ as a function of S/N ratio ($S_{P}/\sigma$)
for all sources in our catalog. The red lines indicate the lower and
upper envelopes. The blue line denotes the $S_{T}=S_{P}$ axis. Sources
(green circles) that lie above the upper envelope are considered to
be resolved.}
\end{figure}

\noindent Using the upper envelope, we find that 4458 of 10091 (i.e.
45 per cent) of the sources in our catalog can be considered extended
(see Fig. \ref{fig:simulations_real}, right panel). These sources
are listed as resolved in the final catalog (Section \ref{sec:source_catalog}).
However, still some objects classified by \texttt{PyBDSF} as made
of multiple components are not identified by this criterion as resolved.
Similarly, point sources could be located above the envelope by chance.

\subsection{Completeness and reliability \label{sec:completeness}}

%1.660, 10.273, 0.024 -->Parameters for 90per envelope<-- [  1.66014526  10.27254556   0.02395406] ii 2753 

\noindent 

The incompleteness in radio surveys is mainly an issue at low S/N
ratios, where a significant fraction of the sources can be missed.
This is consequence of the image noise on the source detection. For
instance, at the detection threshold sources that are located on random
noise peaks are more easily detected than those located on noise dips
\citep{2000A&AS..146...41P}.

\begin{figure*}
\centering{}\centering\includegraphics[scale=0.3,clip=true]{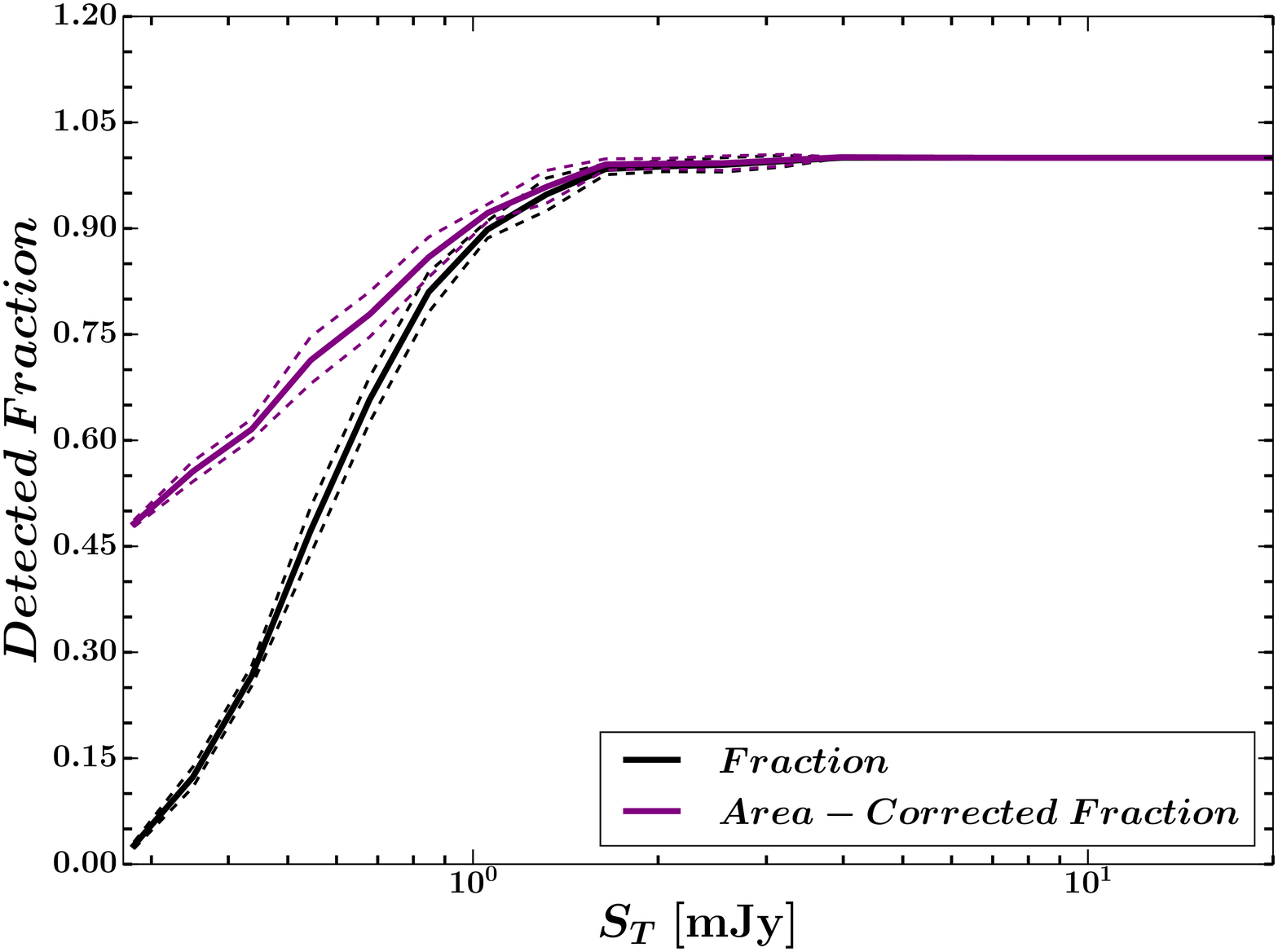}\includegraphics[scale=0.3,clip=true]{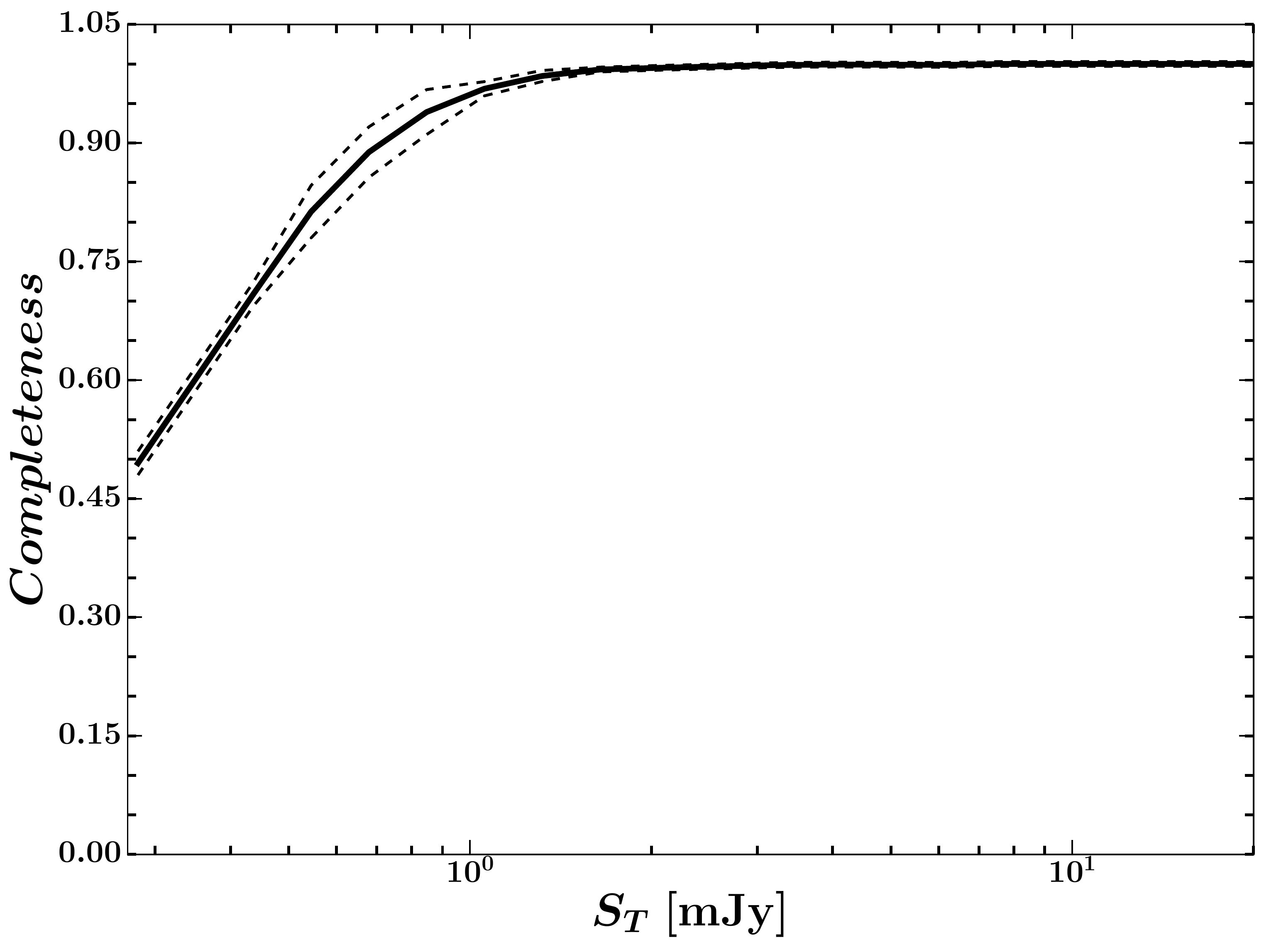}\caption{\label{fig:completeness}Left: The solid black line indicates the
fraction of sources detected in our simulations. The solid purple
line denotes the detected fraction corrected for the visibility area
that is used in the source counts calculation. Right: The completeness
function of our Bo\"otes catalog as a function of flux density. Our
catalog is 95 per cent complete at $S_{\textrm{150MHz}}>1\:\textrm{mJy},$
while the completeness drops to about 80 per cent at $S_{T}\lesssim0.5\:\textrm{mJy}$.
The dashed lines in both plots represent $1\sigma$ errors estimated
using Poisson statistics. }
\end{figure*}

The fraction of sources detected at $5\sigma$ in the mosaic is estimated
through Monte-Carlo (MC) simulations. First, we insert artificial
point sources into the residual map created by \texttt{PyBDSF }(see
Section \texttt{\ref{sec:Section6-1}}). We generate 30 random catalogs
with an artificial source density of at least three times the real
catalog. These artificial sources are placed at random locations in
the residual map. The fluxes are drawn from a power-law distribution
inferred from the real sources, with a range between $0.5\sigma$
and $30\sigma$, where $\sigma=55\:\mu\textrm{Jy/beam}.$ The source
extraction is performed with the same parameters as for the real mosaic.
To obtain a realistic distribution of sources, $40$ per cent of the
objects in our simulated catalogs are taken to be extended. In the
MC simulations, the extended sources are modelled as objects with
a gaussian morphology. Their major axis sizes are drawn randomly from
values between one and two times the synthesized beam size, the minor
axis sizes are chosen to have a fraction between $0.5$ and $1.0$
of the corresponding major axis size, and the position angles are
randomly selected between $0^{\circ}$ and $180^{\circ}$. We determine
the completeness at a specific flux $S_{T}$ by evaluating the integral
distribution of the detected source fraction with total flux $>S_{T}$
. The detected fraction and completeness of our catalog are shown
in Fig. \ref{fig:completeness}. Our results indicate that at $S_{T}>1\:\textrm{mJy},$
our catalog is 95 percent complete, whereas at $S_{T}\lesssim0.5\:\textrm{mJy}$
the completeness drops to about $80\%$. 

\begin{figure*}
\centering{}\centering\includegraphics[scale=0.3,clip=true]{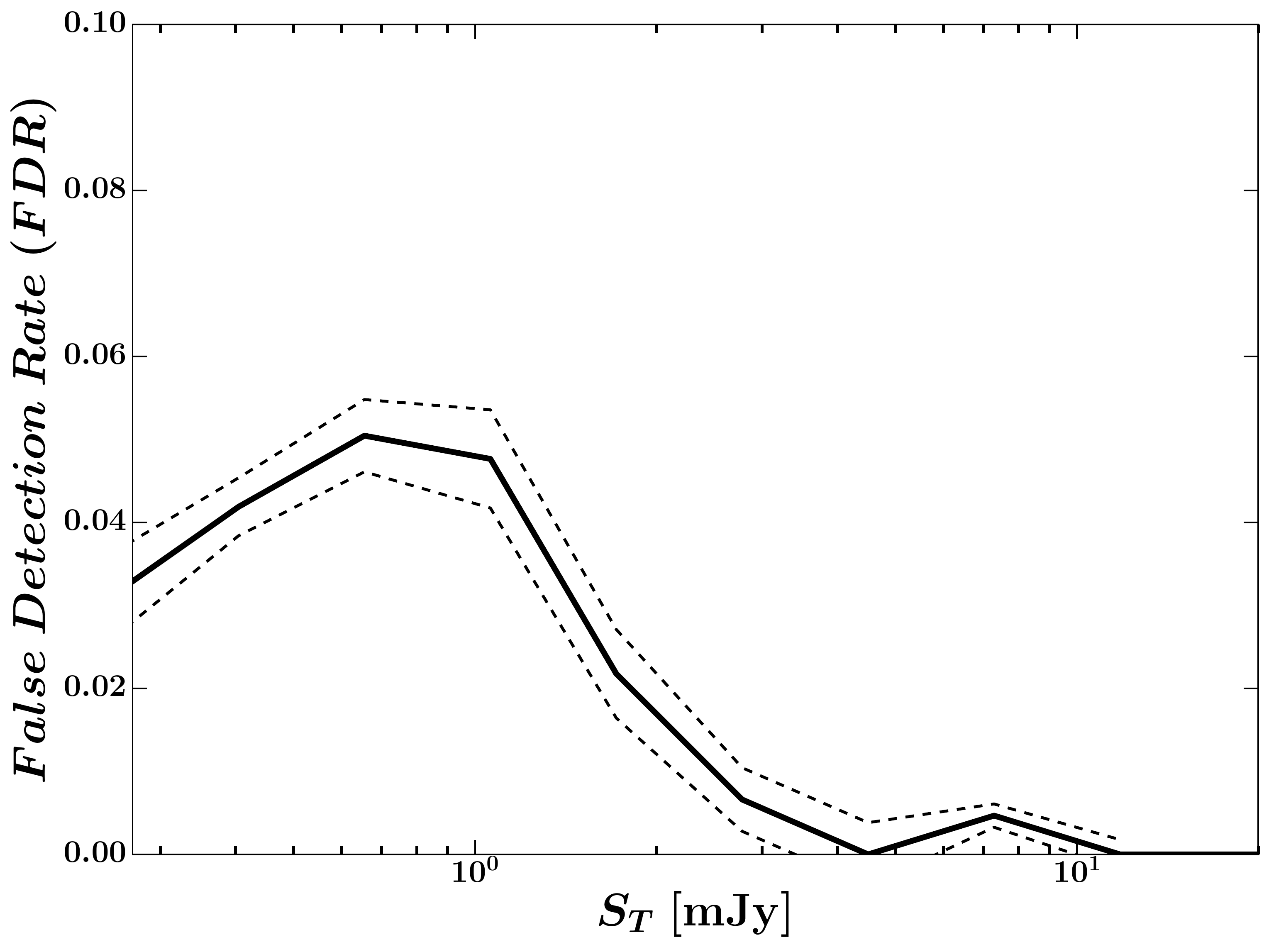}\includegraphics[scale=0.3,clip=true]{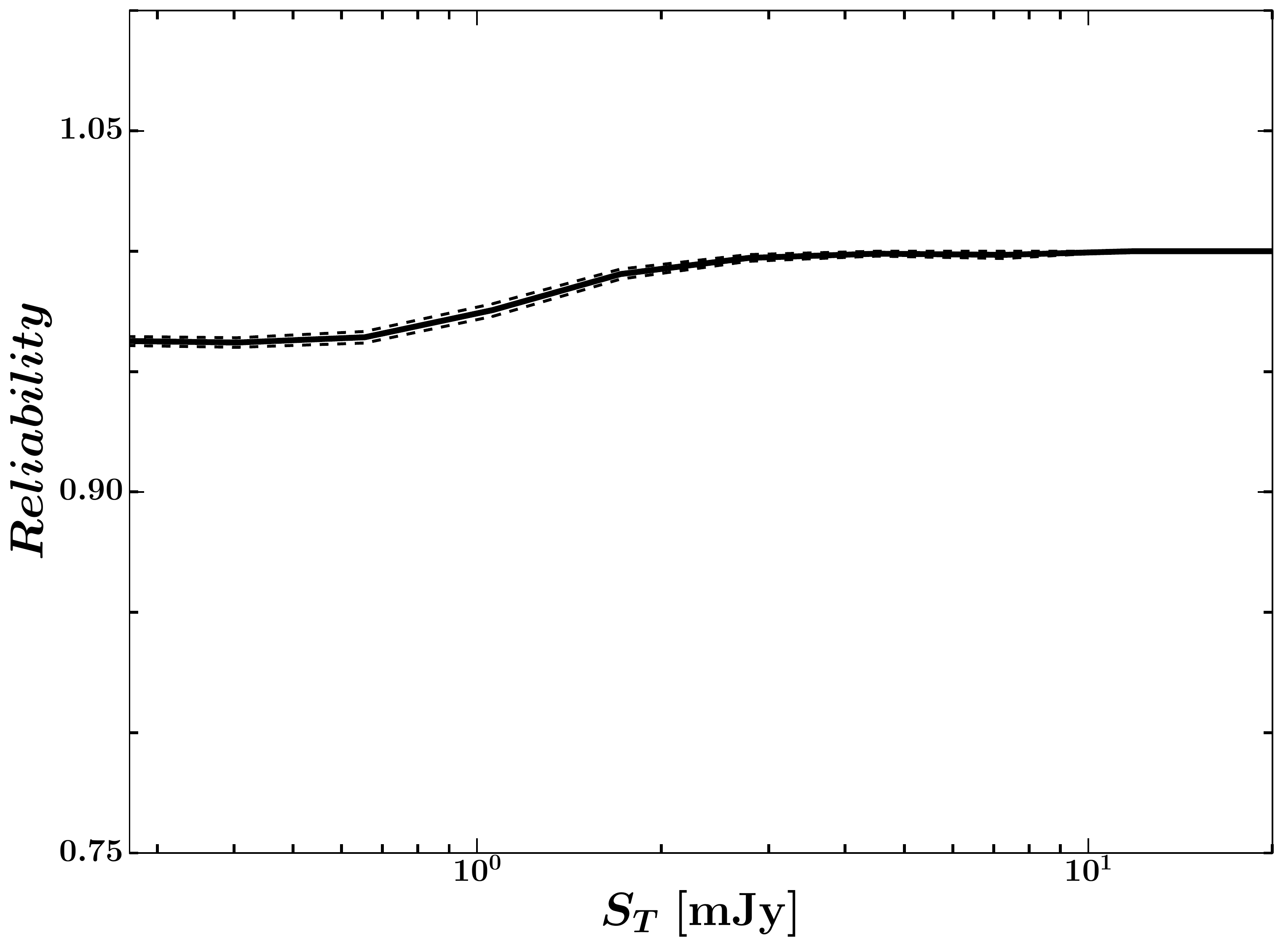}\caption{\label{fig:reliabitiy}Left: False detection rate (FDR) as a function
of flux density. For $S_{T}<1\,\textrm{mJy}$, the FDR is less than
5 per cent, while there are not false detections for $S_{T}>5\,\textrm{mJy}$.
Right: The reliability function of our Bo\"otes catalog as a function
of flux density. The dashed lines in both plots represent $1\sigma$
errors estimated using Poisson statistics. }
\end{figure*}

In our facets, the presence of residual amplitude and phase errors
causes the background noise to deviate from a purely Gaussian distribution.
These noise deviations could be potentially detected by the source-finding
algorithm as real sources. Assuming that the noise deviations can
be equally likely negative or positive and real detections are due
to positive peaks only, we run \texttt{PyBDSF} on the inverted mosaic
as done in Section \texttt{\ref{sec:Section6-1}} to estimate the
false detection rate (FDR) in our survey. This negative mosaic is
created by multiplying all the pixels in the mosaic image by $-1$.
During our tests in the negative mosaic, we discovered that \texttt{PyBDSF}
identifies a large number of artifacts around bright sources as ``real''
sources. This could potentially bias our FDR estimations. Therefore,
we mask the regions around bright sources ($S_{T}>200\textrm{mJy}$)
with circle of radius $25^{\prime\prime}$ to make certain that our
FDR estimations are not dominated by artifacts. Excluding bright sources
does not affect our FDR estimations, as FDR is generally relevant
for fainter sources, whese noise deviations could be detected as real
objects. The FDR is determined from the ratio between the number of
false detections and real detections at a specific flux density bin.
The reliability, $R=1-FDR$ at a given flux density $S$, is estimated
by integrating the FDR over all fluxes $>S$. The FDR and reliability
are plotted as a function of total  flux density in Fig \ref{fig:reliabitiy}. 

\noindent 

\noindent 

\subsection{Source catalog \label{sec:source_catalog}}

The final catalog contains 10091 sources detected above a $5\sigma$
flux density threshold and is made available online\footnote{http://vizier.u-strasbg.fr/viz-bin/VizieR }.
The astrometry, total and peak flux densities in the catalog are corrected
as described in Sections \ref{sec:astrometry_correction}, \ref{sec:smearing_correction},
and \ref{sec:flux_correction}; respectively. The reported flux densities
are on the \citet{2012MNRAS.423L..30S} flux density scale and their
errors have the global uncertainties added in quadrature as described
in Section \ref{sec:flux_correction}. We list a sample from 13 rows
of the published catalog in Table \ref{tab:sample_catalog}, where
the columns are:

$\:$

%\begin{singlespace}
\noindent (1) Source ID

\noindent (2,4) source position (RA, Dec) 

\noindent (3,5) errors in source position

\noindent (6,7) total flux density and error 

\noindent (8,9) peak flux density and error 

\noindent (10) combined bandwidth and time smearing correction factor
for the peak flux density 

\noindent (11) local rms noise

\noindent (12) source type (point source or extended)

\noindent (13) \texttt{PyBDSF }source structure code (S/M)
%\end{singlespace}

$\:$

\noindent Additionally, the catalog contains three flags not shown
in Table \ref{tab:sample_catalog}. These flags follow the naming
convention by \citet{2016MNRAS.460.2385W} as follows:

$\,$

%\begin{singlespace}
\noindent (13) Flag edge, when equals to 1 indicates an object that
is located close to or in a facet edge, which could result in some
flux loss.

\noindent (14) Flag artifact, this flag indicate if an object is a
calibration artifact: a value of ``1'' signifies a source that is
probably an artifact, and ``2'' signifies that is surely an artifact.

\noindent (15) Flag merged, when equal to 1 indicates a large diffuse
source whose separate components are merged into a single one according
to a visual inspection.
%\end{singlespace}

\begin{table*}
\caption{A sample of ten rows from the source catalogue. See Section \ref{sec:source_catalog}
for a description of the columns. \label{tab:sample_catalog}}

\centering{}%
\begin{tabular}{ccccccccccc}
 &  &  &  &  &  &  &  &  &  & \tabularnewline
\hline 
{\tiny{}Source ID } & {\tiny{}RA } & {\tiny{}$\sigma_{\textrm{RA}}$} & {\tiny{}DEC } & {\tiny{}$\sigma_{\textrm{DEC}}$} & {\tiny{}$F_{total}$} & {\tiny{}$F_{peak}$} & {\tiny{}$F_{smear}$} & {\tiny{}$\sigma$} & {\tiny{}Source type} & {\tiny{}PyBDSF code}\tabularnewline
\hline 
{\tiny{}(1) } & {\tiny{}(2) } & {\tiny{}(3) } & {\tiny{}(4) } & {\tiny{}(5) } & {\tiny{}(6,7) } & {\tiny{}(8,9) } & {\tiny{}(10) } & {\tiny{}(11) } & {\tiny{}(12) } & {\tiny{}(13) }\tabularnewline
 & {\tiny{} {[}deg{]}} & {\tiny{}{[}arcsec{]}} & {\tiny{} {[}deg{]}} & {\tiny{}{[}arcsec{]}} & {\tiny{}{[}mJy{]}} & {\tiny{}{[}mJy{]}} &  & {\tiny{}{[}mJy/beam{]}} &  & \tabularnewline
\hline 
{\tiny{}J143941.32+340337.6} & {\tiny{}219.92 } & {\tiny{}0.53 } & {\tiny{}34.06 } & {\tiny{}0.38} & {\tiny{}$0.55\pm0.15$} & {\tiny{}$0.63\pm0.14$} & {\tiny{}1.03} & {\tiny{}0.10} & {\tiny{}P} & {\tiny{}S}\tabularnewline
{\tiny{}J142850.28+323248.3} & {\tiny{}217.21 } & {\tiny{}0.21} & {\tiny{}32.55} & {\tiny{}0.12 } & {\tiny{}$3.63\pm0.58$} & {\tiny{}$2.13\pm0.33$} & {\tiny{}1.05} & {\tiny{}0.10} & {\tiny{}E} & {\tiny{}S}\tabularnewline
{\tiny{}J142948.72+325055.0} & {\tiny{}217.45 } & {\tiny{}0.64} & {\tiny{}32.85} & {\tiny{}0.33} & {\tiny{}$3.3\pm0.58$} & {\tiny{}$2.1\pm0.33$} & {\tiny{}1.04} & {\tiny{}0.07} & {\tiny{}P} & {\tiny{}S}\tabularnewline
{\tiny{}J143105.69+341233.4} & {\tiny{}217.77} & {\tiny{}0.39} & {\tiny{}34.21} & {\tiny{}0.17} & {\tiny{}$0.73\pm0.15$} & {\tiny{}$0.66\pm0.12$} & {\tiny{}1.00} & {\tiny{}0.06} & {\tiny{}P} & {\tiny{}S}\tabularnewline
{\tiny{}J143212.23+340650.8} & {\tiny{}218.05} & {\tiny{}0.60} & {\tiny{}34.11} & {\tiny{}0.21 } & {\tiny{}$0.40\pm0.10$} & {\tiny{}$0.38\pm0.08$} & {\tiny{}1.00} & {\tiny{}0.05} & {\tiny{}P} & {\tiny{}S}\tabularnewline
{\tiny{}J142250.39+334749.8} & {\tiny{}215.71} & {\tiny{}0.35} & {\tiny{}33.80} & {\tiny{}0.19} & {\tiny{}$1.50\pm0.28$} & {\tiny{}$1.19\pm0.21$} & {\tiny{}1.05} & {\tiny{}0.11} & {\tiny{}E} & {\tiny{}S}\tabularnewline
{\tiny{}J144016.42+354346.0} & {\tiny{}220.07} & {\tiny{}0.14} & {\tiny{}35.73} & {\tiny{}0.10 } & {\tiny{}$3.66\pm0.59$} & {\tiny{}$2.98\pm0.46$} & {\tiny{}1.05} & {\tiny{}0.12} & {\tiny{}P} & {\tiny{}S}\tabularnewline
{\tiny{}J143847.45+351001.4} & {\tiny{}219.70} & {\tiny{}0.58} & {\tiny{}35.17} & {\tiny{}0.48} & {\tiny{}$0.62\pm0.14$} & {\tiny{}$0.47\pm0.11$} & {\tiny{}1.03} & {\tiny{}0.07} & {\tiny{}E} & {\tiny{}S}\tabularnewline
{\tiny{}J144101.02+344109.2} & {\tiny{}220.25} & {\tiny{}0.45} & {\tiny{}34.69} & {\tiny{}0.39} & {\tiny{}$0.87\pm0.21$} & {\tiny{}$0.73\pm0.14$} & {\tiny{}1.04} & {\tiny{}0.09} & {\tiny{}P} & {\tiny{}S}\tabularnewline
{\tiny{}J144030.46+354650.3} & {\tiny{}220.13} & {\tiny{}0.53} & {\tiny{}35.78} & {\tiny{}0.41} & {\tiny{}$0.95\pm0.25$} & {\tiny{}$0.86\pm0.18$} & {\tiny{}1.06} & {\tiny{}0.12} & {\tiny{}E} & {\tiny{}S}\tabularnewline
\hline 
\end{tabular}
\end{table*}

\section{Source counts\label{sec:Section7} }

\subsection{Size distribution and resolution bias \label{sec:source_extension} }

%A good knowledge of the angular size distribution of our LOFAR sources is critical for a correct determination of the radio source counts. Particularly, at low-frequencies it is unknown how the angular size distribution varies %towards fainter flux densities. The objects can be more extended at low frequencies, and the size distribution can be different than those estimated in GHz surveys. .

%In this work, we follow the procedure outlined by  with slight modifications.
%First, we measure the maximum extension of a radio source using the total flux  to peak flux  ratio:

%As mentioned in Section , the ratio  can be used to study the extension of a radio source.

\noindent %The ratio  is shown as a function of the local  ratio ( in Fig . The total flux should be equal or larger than the peak flux. In cases when the opposite is true, the image noise is affecting the source fitting on the maps and it is %possible to define a function or envelope that measures the error fluctuations. If we assume that these errors are likely to affect the fits for which , the envelope can be mirrored above the line  and define a boundary for which %sources above it are likely to be resolved. We fit the  envelope with the following function:

%where , , , and  is the local noise level. 
% St/Sp figure caption: Ratio of the total flux  to the peak flux  as a function of the local signal-to-noise  ratio (). The red solid lines represent the lower and upper envelopes that contain  of the unresolved sources. Sources located above the upper %enveloped are considered resolved. Blue points refer to sources that are classified as single by PyBDSM, while green points represent those formed by multiple components.

\begin{figure}[tp]
\centering{}\includegraphics[clip,scale=0.35]{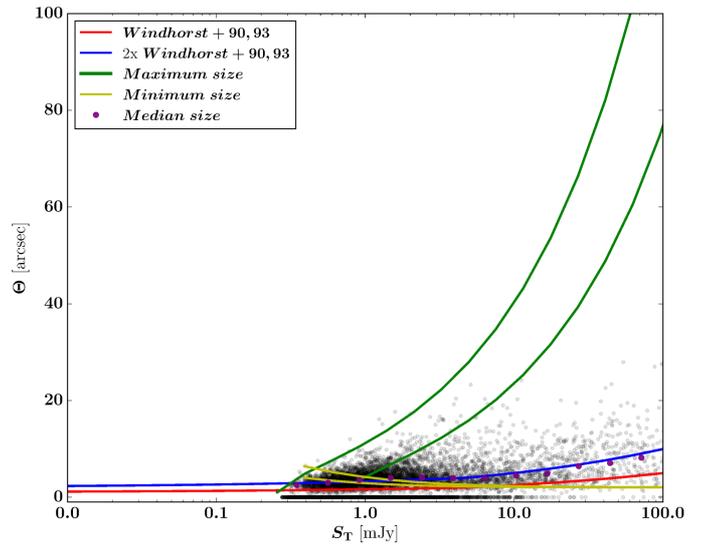}\centering\caption{\label{fig:angulat_size} Angular size (deconvolved geometric mean)
for LOFAR sources as function of their total flux density. The range
of possible values for the maximum and minimum detectable angular
sizes corresponding to the rms range in our mosaic ($55-140\:\mu\textrm{Jy}$)
are indicated by the green and yellow lines, respectively. All unresolved
sources are located in the plane $\Theta=0$, and the median source
sizes for our sample are shown by purple points. The red line indicates
the median of the \citet{1990ASPC...10..389W} functions, the blue
line represents the same function increased by a normalization factor
of 2.}
\end{figure}

\noindent Following \citet{2001A&A...365..392P}, we use estimate
an upper limit $\Theta_{\textrm{lim}}$ for the angular size that
a source of given flux can have before its peak flux falls below our
detection threshold ($5\sigma$). This upper limit is defined as a
function of the total flux density:

\[
\Theta_{\textrm{lim}}=max\left(\Theta_{\textrm{max}},\Theta_{\textrm{min}}\right),
\]

\noindent where $\Theta_{\textrm{max}}$ is obtained utilizing eq.
\ref{eq:St/Sp_vs_angle} and $\Theta_{\textrm{min}}$, the minimum
angular size that is reliably resolved, can be derived combining eqs.
\ref{eq:St/Sp_vs_angle} and \ref{eq:upper_envelope}. The constraint
provided by $\Theta_{\textrm{min}}$ takes into account the finite
size of the synthesized beam and ensures that $\Theta_{\textrm{lim}}$
does not become unphysical ($\Theta_{\textrm{max}}\longrightarrow0$
at low S/N ratios). Sources with sizes $>\Theta_{\textrm{max}}$ will
remain undetected and the resulting catalog will be incomplete, whereas
for sources with sizes $<\Theta_{\textrm{min}}$ the deconvolution
is not reliable. This systematic effect is called resolution bias.
The range of possible values for the $\Theta_{\textrm{max}}$ and
$\Theta_{\textrm{min}}$ according to our rms levels are indicated
by the green and yellow, respectively, shaded lines in Fig \ref{fig:angulat_size}.
To define the rms levels, we consider minimum and maximum noise values
in our map. As shown in Fig. \ref{fig:visibility_area}, 90 per cent
of the total area has approximately $\sigma\lesssim140\:\mu\textrm{Jy}.$
This value can thus be considered as representative of the maximum
noise value. For the minimum noise value, we take the central rms
noise in our map that is about $\sigma\sim55\:\mu\textrm{Jy}$. The
(deconvolved) size distribution of our sources is shown in Fig. \ref{fig:angulat_size}.
As expected our sources tend to be smaller than the maximum allowed
sizes. 

\noindent A good knowledge of the angular size distribution of our
LOFAR sources is critical for a correct determination of the resolution
bias in our survey. Particularly, at low-frequencies the sources can
be more extended, and the size distribution can be different from
that estimated in GHz surveys \citep{2016MNRAS.460.2385W,2016MNRAS.463.2997M}.
In Fig \ref{fig:angulat_size}, we compare the median of the angular
size for our sample (purple points) with the average of the two median
size relations proposed by \citet{1990ASPC...10..389W,1993ApJ...405..498W}
for 1.4GHz surveys:

\[
\Theta_{\textrm{med,1}}=2\,\left(S_{1.4GHz}\right)^{0.3}\,\textrm{arcsec},
\]

%\[
% X(m,n) = \left\{\begin{array}{lr}
% x(n), & \text{for } 0\leq n\leq 1\\
% x(n-1), & \text{for } 0\leq n\leq 1\\
% x(n-1), & \text{for } 0\leq n\leq 1
% \end{array}\right\} = xy
%\]

\[
\Theta_{\textrm{med,2}}=\begin{cases}
2\,\left(S_{1.4GHz}\right)^{0.3}\,\textrm{arcsec} & \:S_{1.4GHz}>1\,\textrm{mJy}\\
2\,\textrm{arcsec} & \:S_{1.4GHz}<1\,\textrm{mJy},
\end{cases}
\]

\noindent after scaling them to 150 MHz using a spectral index of
$\alpha=-0.7$ \citep{2017A&A...602A...1S} (red solid lines). It
is clear that our sources have larger median deconvolved angular sizes
than those predicted by the Windhorst relations. A similar trend was
found by \citet{2016MNRAS.463.2997M} and \citet{2016MNRAS.460.2385W}
in their analysis of LOFAR observations. These authors proposed to
modify the Windhorst relations by increasing the normalization by
factor of 2 (blue solid line) to obtain a better fit to the median
angular sizes for their sources. A close examination to the median
source sizes in our sample indicates that this modification indeed
provides a good fit to our data. Therefore, we employ this relation
to account for the resolution effects in our catalog.

\noindent 

\begin{figure*}[t]
\centering{}\includegraphics[clip,scale=0.55]{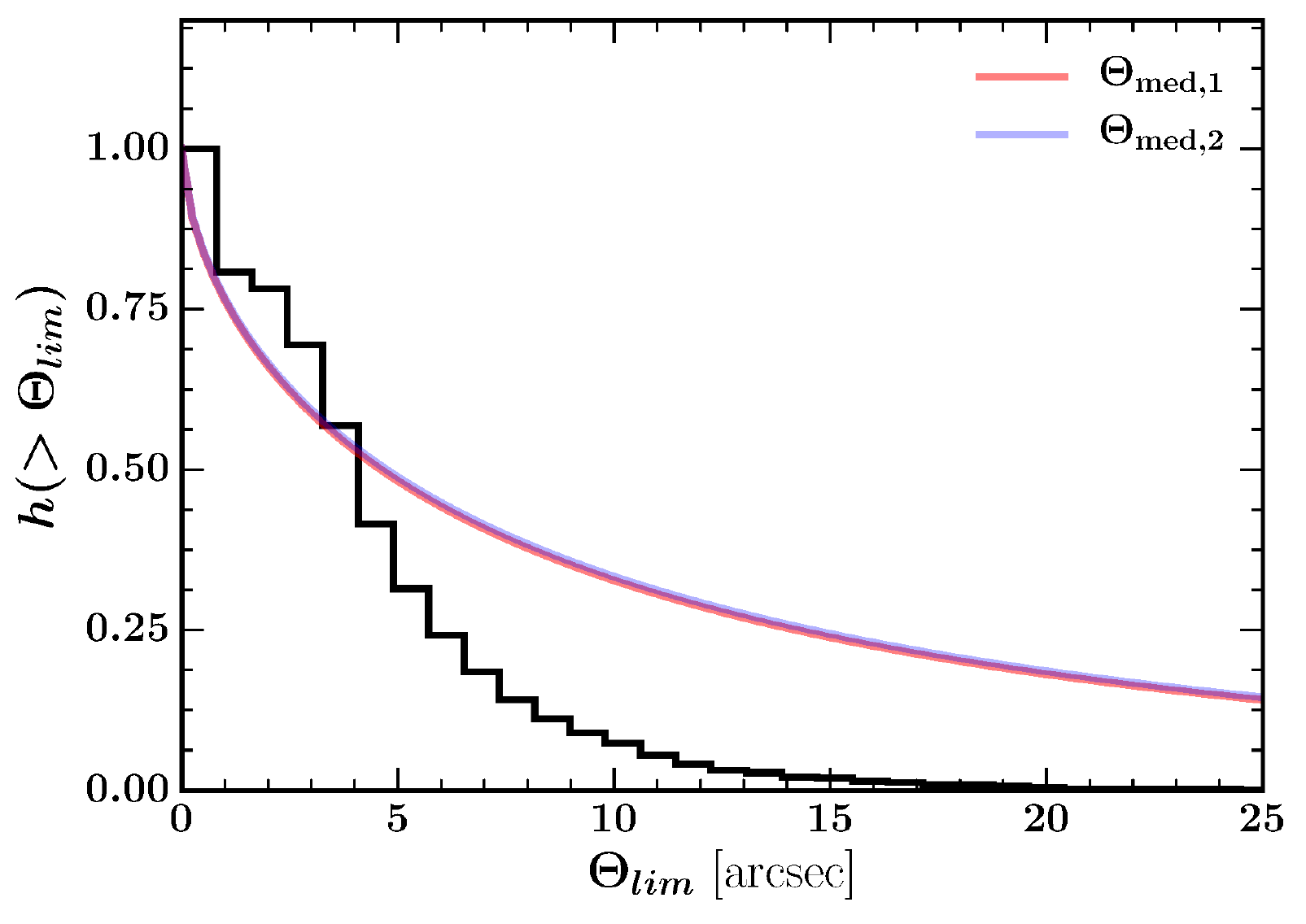}\includegraphics[clip,scale=0.55]{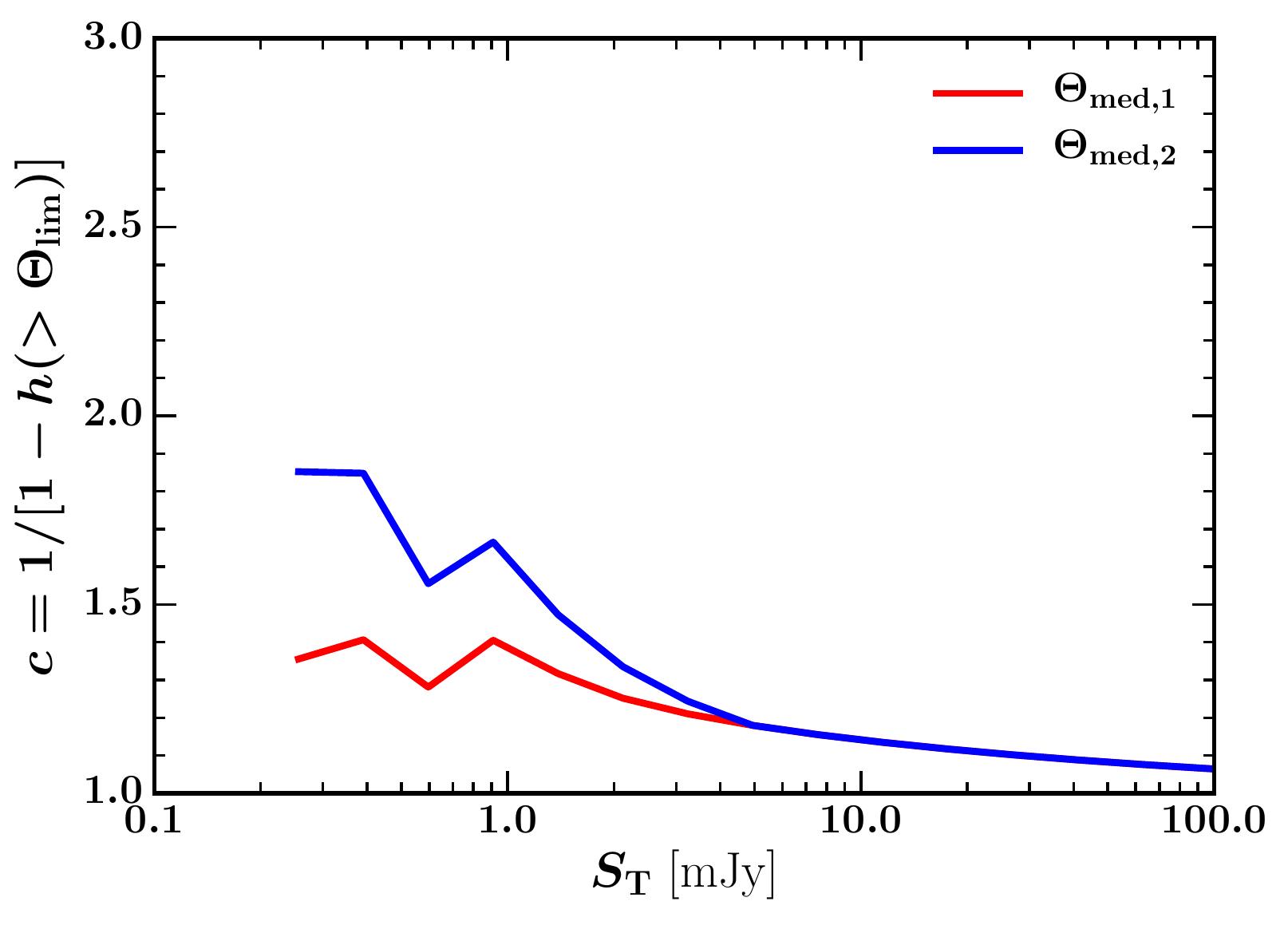}\centering\caption{\label{fig:resolution_bias} Left: Integral size distribution (black
lines) for sources in our catalog with $10\:\textrm{mJy}<S_{150MHz}<25\:\textrm{mJy}$.
The red and blue lines represent the \citet{1990ASPC...10..389W}
and \citet{1993ApJ...405..498W} relations scaled to\textbf{ }150MHz\textbf{
}and increased by a normalization factor of 2 for the corresponding
median angular sizes. Right: The resolution bias $c=1/\left[1-h\left(\Theta_{\textrm{lim}}\right)\right]$
as a function of the total flux density. The color legends are the
same as in the left panel.}
\end{figure*}

\noindent To correct the source counts for the incompleteness due
to the resolution bias we need to determine the \emph{true} integral
angular size distribution of radio sources as a function of the total
flux density. \citet{1990ASPC...10..389W} reported a exponential
form for the true angular size distribution:
\noindent \begin{center}
\begin{equation}
h\left(\Theta_{\textrm{lim}}\right)=\exp\left[\left(b\left(\frac{\Theta_{\textrm{lim}}}{\Theta_{\textrm{med}}}\right)^{a}\right)\right],\label{eq:winhorst_size_dist}
\end{equation}
\par\end{center}

\noindent with $a=-\ln\left(2\right)$ and $b=0.62$. To determine
the unbiased integral size distribution from our sample, we need to
select sources in a total flux density range that is not affected
by the resolution bias. For this purpose, we choose the flux density
range $10\:\textrm{mJy}<S_{T}<25\:\textrm{mJy}$. The reason for choosing
this flux density range is two fold. First, the number of reliably
deconvolved sources in this range is $93\%$, and second to determine
the integral size distribution with a large statistical sample that
is close as possible to our $5\sigma$ detection threshold. In Fig.
\ref{fig:resolution_bias} (lef panel), we compare the integral size
distribution (solid black line) for sources in our catalog with flux
densities in the range $10\:\textrm{mJy}<S_{150MHz}<25\:\textrm{mJy}$
with the 1.4GHz relations proposed by Windhorst scaled to 150 MHz
using a spectral index of $\alpha=-0.7$. We find that the scaled
Windhorst relations are a good represention of the integral size distribution
for $\Theta\lesssim5^{\prime\prime}$ sources, which correspond to
a fraction of 80 per cent in our Bo\"otes catalog. 

\noindent The resolution bias correction is defined as $c=1/\left[1-h\left(\Theta_{\textrm{lim}}\right)\right]$
\citep{2001A&A...365..392P}. Fig. \ref{fig:resolution_bias} (right
panel) shows the resolution bias correction as a function of the total
flux density for the scaled Windhorst relations and the integral size
distribution determined for our sample. We use the average of the
Windhorst relations to apply the resolution bias correction to our
catalog. Additionally, a 10 per cent uncertainty is added in quadrature
to the errors in the source counts following \citet{1990ASPC...10..389W}.

\noindent 

\subsection{Visibility area}

%The variation in the noise across our map is dictated by the shape of the primary beam shape

The varying noise present in our mosaic implies that objects with
different flux densities are not distributed uniformly in the region
surveyed. Thus, the contribution of each object to the source counts
is weighted by the reciprocal of its visibility area (i.e. the fraction
of the total area in which the source can be detected), as derived
in Section \ref{sec:Section6-1}. This correction allows us to account
for different visibility areas within the same flux density bin.

\subsection{Completeness and reliability}

As can be seen in Fig. \ref{sec:completeness}, the fraction of detected
sources decreases towards fainter flux densities. Thus, a correction
factor that accounts for the missed objects is required when calculating
the source counts. For this purpose, we employ the detected fraction
corrected for the visibility area (see Fig. \ref{fig:completeness})
to account for the incompleteness in our source counts. Furthermore,
we apply a factor to account for the reliability using the FDR derived
in Section \ref{sec:completeness}.

%The completeness and reliability of our Botes catalog We use the results from Section  to correct for completeness and reliability in our catalog. However, because the visibility area is used for each source, we employ the %corrected-area detected fraction (see Fig. ) to correct for incompleteness effects in our catalog. Usin

\subsection{Multiple-component sources}

%As can be seen in Fig. , many sources in our survey are resolved out and split out into multiple-components that are likely physically related. 

In Section \ref{sec:Section6-1}, we carried out a visual inspection
to identify resolved sources that have been misclassified into different
single components by our source extraction software. However, for
sources that are resolved out and split out into multiple-components
and do not show signs of physical connection, establishing that their
components are part of a same source is not trivial. Consequently,
these components are still listed as separate sources in our catalog.
This must be taken into account when computing the source counts to
ensure these multi-component sources are only counted once. For this
purpose, we employ the algorithm by \citet{1998MNRAS.300..257M},
to identity the missed double sources in our catalog. First, the separation
between a component and its nearest neighbor, and the total flux density
of the two components are compared. The components are considered
as part of a double source if their flux ratio $f$ is in the range
$0.25\leq f\leq4$, and satisfies the separation criterion scaled
to 150MHz using a spectral index of $\alpha=-0.7$:
\noindent \begin{center}
$\Theta_{0}<100\,\sqrt{\left(\frac{S_{T}}{20}\right)},$
\par\end{center}

\noindent where $\Theta_{0}$ is in arcseconds and $S_{T}$ is the
summed flux of the two components, otherwise the components are considered
independent single sources . We identify 633 sources (i.e. 6 per cent
of the catalog) as doubles following the \citet{1998MNRAS.300..257M}
criterion.

\noindent %For comparison, only 246 objects are found within the 20 per cent power point of the primary beam that could be classified as double sources using the same criterium.

\noindent %We identify 648 sources (i.e. 6 percent of the catalog) as doubles following the  criteria for the source counts calculation.

\subsection{Differential source counts}

The normalized 150Hz differential radio-source counts derived from
our LOFAR Bo\"otes observations between our $5\sigma$ flux density
threshold of $275\,\mu\textrm{Jy}$ and $3\,\textrm{Jy}$ are shown
in Fig. \ref{fig:source_counts}. Vertical error bars indicate the
uncertainties obtained by propagating the errors on the correction
factors to the $\sqrt{n}$ Poissonian errors \citep{1986ApJ...303..336G}
from the raw counts. Horizontal error bars denote the flux bins width. 

%The source counts by  are from the TGSS survey that covers  in the northern hemisphere with GMRT, and those by  are from a deep MWA pointing over .  We find that the  scatter of the circular sector source counts is  dex. 

\noindent For comparison purposes, previous 150 MHz source counts
by \citet{2017AA...598A..78I} and \citet{2016MNRAS.459.3314F}, as
well as the Bo\"otes counts obtained by \citet{2016MNRAS.460.2385W}
are shown in Fig. \ref{fig:source_counts}. Additionally, we show
previous results from deep fields at 1.4GHz \citep{2015MNRAS.452.1263P},
3GHz \citep{2017A&A...602A...1S} and the compilation by \citet{2010A&ARv..18....1D}
scaled to 150 MHz using a spectral index of $\alpha=-0.7$ \citep{2017A&A...602A...1S}.

\noindent Our source counts are in fairly good agreement with previous
low- and high- frequency surveys. At $S_{\textrm{150MHz}}>1\,\textrm{mJy}$,
there is a very good consistency for the source counts derived from
the various surveys. The situation is different for the fainter flux
bins ($S_{\textrm{150MHz}}<1\,\textrm{mJy}$), where there is a large
dispersion between the results from the literature. In the range $S_{150MHz}\leq1.0\:\textrm{mJy},$
our source counts are consistent with those derived by \citet{2016MNRAS.460.2385W},
and also they closely follow the counts reported by \citet{2017A&A...602A...1S}.
In the flux density bins $S_{150MHz}\leq0.4\:\textrm{mJy},$ the drop
in the source counts may be the result of residual incompleteness.
Our data confirms the change in the slope at sub-mJy flux densities
previously reported in the literature by high- \citep{1988AA...195...21K,1998MNRAS.296..839H,2015MNRAS.452.1263P}
and low- \citep{2016MNRAS.460.2385W,2016MNRAS.463.2997M} frequency
surveys. This change can be associated to the increasing contribution
of SF galaxies and radio-quiet AGNs at the faintest flux density bins
\citep{2008ApJS..177...14S,2009ApJ...694..235P,2011ApJ...740...20P,2017A&A...602A...2S}. 

\subsection{Cosmic variance}

\begin{figure}[tp]
\centering{}\includegraphics[clip,scale=0.32]{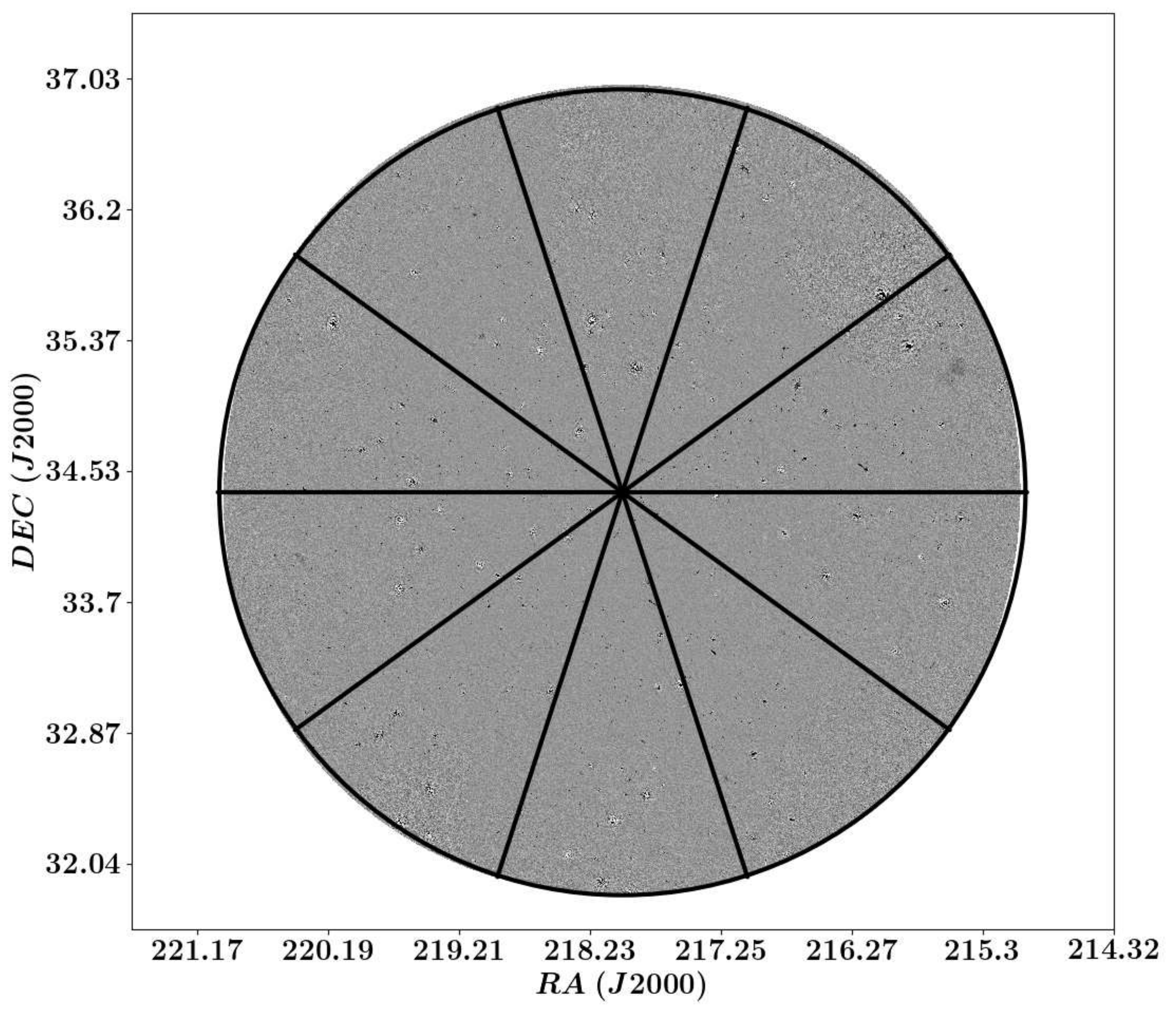}\centering\caption{\label{fig:circular_sectors} The spatial distribution of the circular
sectors in the Bo\"otes mosaic used to test the effect of cosmic
variance in our source counts. Each circular sector has an approximate
area of $2\:\textrm{deg}^{2}$.}
\end{figure}

The differences between source counts at flux densities $<1.0\:\textrm{mJy}$
for multiple independent fields are generally larger than predicted
from their Poissonian fluctuations \citep{2007ASPC..380..189C}. These
differences may result from either systematics uncertainties such
as the calibration accuracy, primary beam correction, and bandwidth
smearing, or different resolution bias corrections adopted in the
literature, or cosmic variance introduced by the large scale structure.
The combination of large area coverage and high sensitivity of our
Bo\"otes observations offers an excellent opportunity to investigate
the effect of cosmic variance in the source counts from different
extragalactic fields. For this purpose, we divide the $20\:\textrm{deg}^{2}$
Bo\"otes mosaic into 10 non-overlapping circular sectors, each one
with an approximate area of $2\:\textrm{deg}^{2}$ and on average
containing more than 900 sources. Fig. \ref{fig:circular_sectors}
shows the spatial distribution of the circular sectors in the Bo\"otes
mosaic.

\noindent 

\begin{figure*}[tp]
\centering{}\includegraphics[clip,scale=0.65]{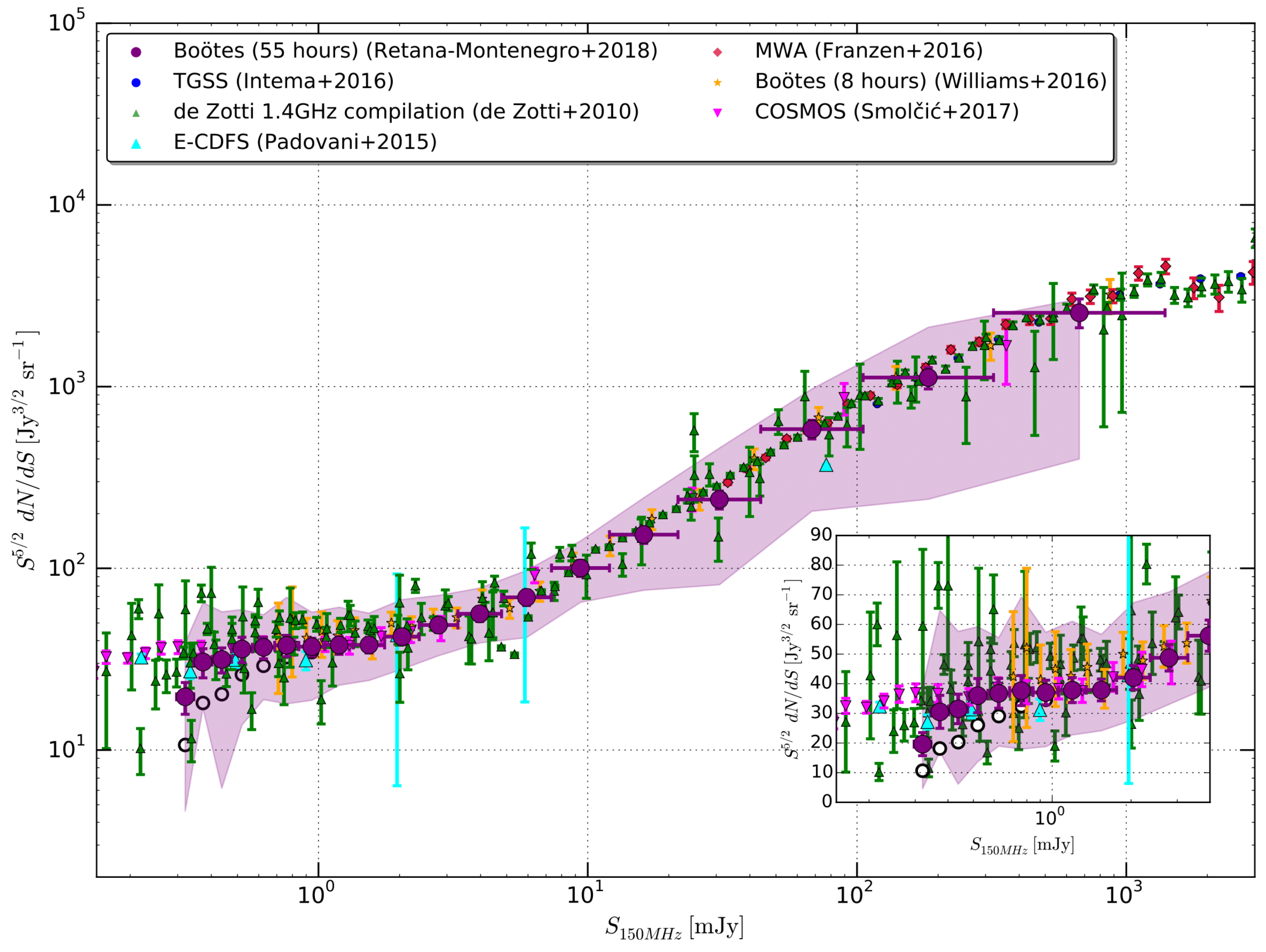}\centering\caption{\label{fig:source_counts} Normalized 150Hz differential radio-source
counts derived from our LOFAR Bo\"otes observations between $275\,\mu\textrm{Jy}$
and $2$ Jy (purple points). Vertical error bars are calculated assuming
Poissonian statistics and horizontal error bars denote the flux bins
width. Open black circles show the counts uncorrected for completeness
and reliability. The purple shaded area displays the $1\sigma$ range
of source counts derived from 10 non-overlapping circular sectors.
For comparison, we overplot the source counts from recent deep and
wide low-frequency surveys \citep{2016MNRAS.459.3314F,2017AA...598A..78I},
as well the source counts derived by \citet{2016MNRAS.460.2385W}
in the Bo\"otes field. In addition, the results of previous deep
surveys carried out at 1.4GHz \citep{2010A&ARv..18....1D,2015MNRAS.452.1263P};
and 3GHz \citep{2017A&A...602A...1S} are scaled to 150 MHz using
a spectral index of $\alpha=-0.7$ \citep{2017A&A...602A...1S}. The
inset shows the source counts in the range $0.080\:\textrm{mJy}\leq S_{\textrm{150MHz}}\leq4\:\textrm{mJy}$.}
\end{figure*}

The purple shaded region in Fig. \ref{fig:source_counts} shows the
$1\sigma$ scatter due to cosmic variance in our source counts. The
counts in each circular sector are computed in the same way as done
for the entire mosaic. The comparison of the shaded region with the
counts derived from deep observations scaled to 150MHz suggests that
the $1\sigma$ scatter due to cosmic variance is larger than the Poissonian
errors of the source counts, and it may explain the dispersion from
previously reported depth source counts at flux densities $S<1\,\textrm{mJy}$.
This confirms the results of \citet{2013MNRAS.432.2625H} who reached
a similar conclusion by comparing the scatter of observed source counts
with that of matched samples from the S3-SEX simulation by \citet{2008MNRAS.388.1335W}.

%We find that the scatter induced by cosmic variance, at least down to , decreases towards the low flux density bins as the number of radio sources increases.

%ndicates that the scatter from previous measurements are dominated by cosmic variance, and may explain the scatter from previous depth counts measurements at flux densities .

\noindent %Such dispersion is often attributed to the different resolution bias corrections adopted in the literature and cosmic variance . The twenty square degrees of our Botes mosaic minimizes the effect of cosmic variance in our derived %source counts.

%In the range  our source counts are slightly higher than those derived by . This trend for the LOFAR source counts is consistent with the results from .

\noindent %Only sources within the 20 per cent power point of the primary beam are considered for the source counts estimation. This large of area seven square degrees minimizes the effect of cosmic variance in our derived source counts. 

% In this range, the LOFAR source counts are higher by a factor of when compared to other results scaled to 150MHz. 

\noindent %Several deep high-frequency surveys have found a strong evolution on faint radio source counts due to the from startburst radio activity.
%Several reasons can explain these differences: cosmic variance, incorrect completeness corrections, different schemes for resolution bias, and clustering. 
%We notice that according to the ATESP determination the upturn in the source counts should show up at lower fluxes (S < 1 mJy) than indicated by the Windhorst et al. (1990) fit (S  5 mJy). The ATESP counts are in very good %agreement with the FIRST counts (White et al. 1997,
%Our counts are in good agreement, over the entire flux range sampled by our data (0.0810 mJy), with the best fit to earlier surveys (Katgert et al. 1988). It is interesting to note that, because of the relatively good statistics of %our data points over about two orders of magnitude in flux, our data clearly show the change in slope of the differential counts, occurring below 1 mJy. Fitting the VLA-VDF differential and integral counts with two power laws we %obtain, for S < 0.6 mJy: triangles in Fig. 6), which are the most accurate available today over the flux range 20 mJy. Our survey, on the other hand, provides the best determination of the counts at fainter fluxes (0.7 < S < 2 %mJy), where the FIRST be- comes incomplete. The ATESP counts can thus provide an useful observational constraint on the evolutionary models for the mJy and sub-mJy populations. 

\noindent 

\section{Conclusions\label{sec:Section8}}

We have presented deep LOFAR observations at 150 MHz. These observations
cover the entire Bo\"otes field down to an rms noise level of $\sim55\:\mu\textrm{Jy}/\textrm{beam}$
in the inner region, with a synthesized beam of $3.98^{''}\times6.45^{''}$.
Our radio catalog contains 10091 entries above the $5\sigma$ detection
over an area of $20\:\textrm{deg}^{2}$. We investigated the astrometry,
flux scale accuracy and other systematics in our source catalog. Our
radio source counts are in agreement with those derived from deep
high-frequency surveys and recent low-frequency observations. Additionally,
we confirm the sharp change in the counts slope at sub-mJy flux densities.
The combination of large area coverage and high sensitivity of our
Bo\"otes observations suggests that the $1\sigma$ scatter due to
cosmic variance is larger than the Poissonian errors of the source
counts, and it may explain the dispersion from previously reported
depth source counts at flux densities $S<1\,\textrm{mJy}$.

\noindent Our LOFAR observations combined with the Bo\"otes ancillary
data will allow us to perform a photometric identification of most
of the newly detected radio sources in the catalog, including rare
objects such as high-z quasars \citep{10.3389/fspas.2018.00005}.
Future spectroscopic observations will provide an unique opportunity
to study the nature of these faint low-frequency radio sources.

\begin{acknowledgements}
ERM acknowledges financial support from NWO Top project No. 614.001.006. HJAR and RJvW acknowledge support from the ERC Advanced Investigator program NewClusters 321271. PNB is grateful for support from the UK STFC via grant ST/M001229/1. \\
LOFAR, the Low Frequency Array designed and constructed by ASTRON, has facilities in several countries, that are owned by various parties (each with their own funding sources), and that are collectively operated by the International LOFAR Telescope (ILT) foundation under a joint scientific policy. The Open University is incorporated by Royal Charter (RC 000391), an exempt charity in England \& Wales and a charity registered in Scotland (SC 038302). The Open University is authorized and regulated by the Financial Conduct Authority.
\end{acknowledgements}

\bibliographystyle{aa}
\addcontentsline{toc}{section}{\refname}\bibliography{my_bib}

\begin{thebibliography}{71}
\expandafter\ifx\csname natexlab\endcsname\relax\def\natexlab#1{#1}\fi

\bibitem[{{Ashby} {et~al.}(2009){Ashby}, {Stern}, {Brodwin}, {Griffith},
  {Eisenhardt}, {Koz{\l}owski}, {Kochanek}, {Bock}, {Borys}, {Brand}, {Brown},
  {Cool}, {Cooray}, {Croft}, {Dey}, {Eisenstein}, {Gonzalez}, {Gorjian},
  {Grogin}, {Ivison}, {Jacob}, {Jannuzi}, {Mainzer}, {Moustakas},
  {R{\"o}ttgering}, {Seymour}, {Smith}, {Stanford}, {Stauffer}, {Sullivan},
  {van Breugel}, {Willner}, \& {Wright}}]{2009ApJ...701..428A}
{Ashby}, M.~L.~N., {Stern}, D., {Brodwin}, M., {et~al.} 2009, \apj, 701, 428

\bibitem[{{Autry} {et~al.}(2003){Autry}, {Probst}, {Starr}, {Abdel-Gawad},
  {Blakley}, {Daly}, {Dominguez}, {Hileman}, {Liang}, {Pearson}, {Shaw}, \&
  {Tody}}]{2003SPIE.4841..525A}
{Autry}, R.~G., {Probst}, R.~G., {Starr}, B.~M., {et~al.} 2003, in \procspie,
  Vol. 4841, Instrument Design and Performance for Optical/Infrared
  Ground-based Telescopes, ed. M.~{Iye} \& A.~F.~M. {Moorwood}, 525--539

\bibitem[{{Becker} {et~al.}(1995){Becker}, {White}, \&
  {Helfand}}]{1995ApJ...450..559B}
{Becker}, R.~H., {White}, R.~L., \& {Helfand}, D.~J. 1995, \apj, 450, 559

\bibitem[{{Bian} {et~al.}(2013){Bian}, {Fan}, {Jiang}, {McGreer}, {Dey},
  {Green}, {Maiolino}, {Walter}, {Lee}, \& {Dav{\'e}}}]{2013ApJ...774...28B}
{Bian}, F., {Fan}, X., {Jiang}, L., {et~al.} 2013, \apj, 774, 28

\bibitem[{{Bridle} \& {Schwab}(1999)}]{1999ASPC..180..371B}
{Bridle}, A.~H. \& {Schwab}, F.~R. 1999, in Astronomical Society of the Pacific
  Conference Series, Vol. 180, Synthesis Imaging in Radio Astronomy II, ed.
  G.~B. {Taylor}, C.~L. {Carilli}, \& R.~A. {Perley}, 371

\bibitem[{{Cohen} {et~al.}(2007){Cohen}, {Lane}, {Cotton}, {Kassim}, {Lazio},
  {Perley}, {Condon}, \& {Erickson}}]{2007AJ....134.1245C}
{Cohen}, A.~S., {Lane}, W.~M., {Cotton}, W.~D., {et~al.} 2007, \aj, 134, 1245

\bibitem[{{Condon}(1992)}]{1992ARAA..30..575C}
{Condon}, J.~J. 1992, \araa, 30, 575

\bibitem[{{Condon}(2007)}]{2007ASPC..380..189C}
{Condon}, J.~J. 2007, in Astronomical Society of the Pacific Conference Series,
  Vol. 380, Deepest Astronomical Surveys, ed. J.~{Afonso}, H.~C. {Ferguson},
  B.~{Mobasher}, \& R.~{Norris}, 189

\bibitem[{{Condon} {et~al.}(2012){Condon}, {Cotton}, {Fomalont}, {Kellermann},
  {Miller}, {Perley}, {Scott}, {Vernstrom}, \& {Wall}}]{2012ApJ...758...23C}
{Condon}, J.~J., {Cotton}, W.~D., {Fomalont}, E.~B., {et~al.} 2012, \apj, 758,
  23

\bibitem[{{Condon} {et~al.}(1998){Condon}, {Cotton}, {Greisen}, {Yin},
  {Perley}, {Taylor}, \& {Broderick}}]{1998AJ....115.1693C}
{Condon}, J.~J., {Cotton}, W.~D., {Greisen}, E.~W., {et~al.} 1998, \aj, 115,
  1693

\bibitem[{{Cool}(2007)}]{2007ApJS..169...21C}
{Cool}, R.~J. 2007, \apjs, 169, 21

\bibitem[{{Cotton} {et~al.}(2004){Cotton}, {Condon}, {Perley}, {Kassim},
  {Lazio}, {Cohen}, {Lane}, \& {Erickson}}]{2004SPIE.5489..180C}
{Cotton}, W.~D., {Condon}, J.~J., {Perley}, R.~A., {et~al.} 2004, in \procspie,
  Vol. 5489, Ground-based Telescopes, ed. J.~M. {Oschmann}, Jr., 180--189

\bibitem[{{de Vries} {et~al.}(2002){de Vries}, {Morganti}, {R{\"o}ttgering},
  {Vermeulen}, {van Breugel}, {Rengelink}, \& {Jarvis}}]{2002AJ....123.1784D}
{de Vries}, W.~H., {Morganti}, R., {R{\"o}ttgering}, H.~J.~A., {et~al.} 2002,
  \aj, 123, 1784

\bibitem[{{de Zotti} {et~al.}(2010){de Zotti}, {Massardi}, {Negrello}, \&
  {Wall}}]{2010A&ARv..18....1D}
{de Zotti}, G., {Massardi}, M., {Negrello}, M., \& {Wall}, J. 2010, \aapr, 18,
  1

\bibitem[{{Franzen} {et~al.}(2016){Franzen}, {Jackson}, {Offringa}, {Ekers},
  {Wayth}, {Bernardi}, {Bowman}, {Briggs}, {Cappallo}, {Deshpande}, {Gaensler},
  {Greenhill}, {Hazelton}, {Johnston-Hollitt}, {Kaplan}, {Lonsdale},
  {McWhirter}, {Mitchell}, {Morales}, {Morgan}, {Morgan}, {Oberoi}, {Ord},
  {Prabu}, {Seymour}, {Shankar}, {Srivani}, {Subrahmanyan}, {Tingay}, {Trott},
  {Webster}, {Williams}, \& {Williams}}]{2016MNRAS.459.3314F}
{Franzen}, T.~M.~O., {Jackson}, C.~A., {Offringa}, A.~R., {et~al.} 2016,
  \mnras, 459, 3314

\bibitem[{{Gehrels}(1986)}]{1986ApJ...303..336G}
{Gehrels}, N. 1986, \apj, 303, 336

\bibitem[{{Hales} {et~al.}(1988){Hales}, {Baldwin}, \&
  {Warner}}]{1988MNRAS.234..919H}
{Hales}, S.~E.~G., {Baldwin}, J.~E., \& {Warner}, P.~J. 1988, \mnras, 234, 919

\bibitem[{{Heald} {et~al.}(2015){Heald}, {Pizzo}, {Orr{\'u}}, {Breton},
  {Carbone}, {Ferrari}, {Hardcastle}, {Jurusik}, {Macario}, {Mulcahy},
  {Rafferty}, {Asgekar}, {Brentjens}, {Fallows}, {Frieswijk}, {Toribio},
  {Adebahr}, {Arts}, {Bell}, {Bonafede}, {Bray}, {Broderick}, {Cantwell},
  {Carroll}, {Cendes}, {Clarke}, {Croston}, {Daiboo}, {de Gasperin}, {Gregson},
  {Harwood}, {Hassall}, {Heesen}, {Horneffer}, {van der Horst}, {Iacobelli},
  {Jeli{\'c}}, {Jones}, {Kant}, {Kokotanekov}, {Martin}, {McKean}, {Morabito},
  {Nikiel-Wroczy{\'n}ski}, {Offringa}, {Pandey}, {Pandey-Pommier}, {Pietka},
  {Pratley}, {Riseley}, {Rowlinson}, {Sabater}, {Scaife}, {Scheers},
  {Sendlinger}, {Shulevski}, {Sipior}, {Sobey}, {Stewart}, {Stroe}, {Swinbank},
  {Tasse}, {Tr{\"u}stedt}, {Varenius}, {van Velzen}, {Vilchez}, {van Weeren},
  {Wijnholds}, {Williams}, {de Bruyn}, {Nijboer}, {Wise}, {Alexov}, {Anderson},
  {Avruch}, {Beck}, {Bell}, {van Bemmel}, {Bentum}, {Bernardi}, {Best},
  {Breitling}, {Brouw}, {Br{\"u}ggen}, {Butcher}, {Ciardi}, {Conway}, {de
  Geus}, {de Jong}, {de Vos}, {Deller}, {Dettmar}, {Duscha}, {Eisl{\"o}ffel},
  {Engels}, {Falcke}, {Fender}, {Garrett}, {Grie{\ss}meier}, {Gunst},
  {Hamaker}, {Hessels}, {Hoeft}, {H{\"o}randel}, {Holties}, {Intema},
  {Jackson}, {J{\"u}tte}, {Karastergiou}, {Klijn}, {Kondratiev}, {Koopmans},
  {Kuniyoshi}, {Kuper}, {Law}, {van Leeuwen}, {Loose}, {Maat}, {Markoff},
  {McFadden}, {McKay-Bukowski}, {Mevius}, {Miller-Jones}, {Morganti}, {Munk},
  {Nelles}, {Noordam}, {Norden}, {Paas}, {Polatidis}, {Reich}, {Renting},
  {R{\"o}ttgering}, {Schoenmakers}, {Schwarz}, {Sluman}, {Smirnov}, {Stappers},
  {Steinmetz}, {Tagger}, {Tang}, {ter Veen}, {Thoudam}, {Vermeulen}, {Vocks},
  {Vogt}, {Wijers}, {Wucknitz}, {Yatawatta}, \& {Zarka}}]{2015A&A...582A.123H}
{Heald}, G.~H., {Pizzo}, R.~F., {Orr{\'u}}, E., {et~al.} 2015, \aap, 582, A123

\bibitem[{{Heywood} {et~al.}(2013){Heywood}, {Jarvis}, \&
  {Condon}}]{2013MNRAS.432.2625H}
{Heywood}, I., {Jarvis}, M.~J., \& {Condon}, J.~J. 2013, \mnras, 432, 2625

\bibitem[{{Hopkins} {et~al.}(1998){Hopkins}, {Mobasher}, {Cram}, \&
  {Rowan-Robinson}}]{1998MNRAS.296..839H}
{Hopkins}, A.~M., {Mobasher}, B., {Cram}, L., \& {Rowan-Robinson}, M. 1998,
  \mnras, 296, 839

\bibitem[{{Intema} {et~al.}(2017){Intema}, {Jagannathan}, {Mooley}, \&
  {Frail}}]{2017AA...598A..78I}
{Intema}, H.~T., {Jagannathan}, P., {Mooley}, K.~P., \& {Frail}, D.~A. 2017,
  \aap, 598, A78

\bibitem[{{Intema} {et~al.}(2009){Intema}, {van der Tol}, {Cotton}, {Cohen},
  {van Bemmel}, \& {R{\"o}ttgering}}]{2009A&A...501.1185I}
{Intema}, H.~T., {van der Tol}, S., {Cotton}, W.~D., {et~al.} 2009, \aap, 501,
  1185

\bibitem[{{Jannuzi} {et~al.}(2010){Jannuzi}, {Weiner}, {Block}, {Borys},
  {Eisenstein}, {Kochanek}, {Rieke}, {Rieke}, {Armus}, {Brodwin}, {Brown},
  {Cool}, {Desai}, {Dey}, {Dickinson}, {Dole}, {Herrera}, {Le Floc'h},
  {Morrison}, {Papovich}, {P{\'e}rez-Gonz{\'a}lez}, {Stern}, {Rujopakarn}, \&
  {Zehavi}}]{2010AAS...21547001J}
{Jannuzi}, B., {Weiner}, B., {Block}, M., {et~al.} 2010, in Bulletin of the
  American Astronomical Society, Vol.~42, American Astronomical Society Meeting
  Abstracts \#215, 513

\bibitem[{{Jannuzi} \& {Dey}(1999)}]{1999ASPC..191..111J}
{Jannuzi}, B.~T. \& {Dey}, A. 1999, in Astronomical Society of the Pacific
  Conference Series, Vol. 191, Photometric Redshifts and the Detection of High
  Redshift Galaxies, ed. R.~{Weymann}, L.~{Storrie-Lombardi}, M.~{Sawicki}, \&
  R.~{Brunner}, 111

\bibitem[{{Katgert} {et~al.}(1988){Katgert}, {Oort}, \&
  {Windhorst}}]{1988AA...195...21K}
{Katgert}, P., {Oort}, M.~J.~A., \& {Windhorst}, R.~A. 1988, \aap, 195, 21

\bibitem[{{Kazemi} {et~al.}(2011){Kazemi}, {Yatawatta}, {Zaroubi},
  {Lampropoulos}, {de Bruyn}, {Koopmans}, \& {Noordam}}]{2011MNRAS.414.1656K}
{Kazemi}, S., {Yatawatta}, S., {Zaroubi}, S., {et~al.} 2011, \mnras, 414, 1656

\bibitem[{{Kellermann} {et~al.}(1986){Kellermann}, {Fomalont}, {Weistrop}, \&
  {Wall}}]{1986HiA.....7..367K}
{Kellermann}, K.~I., {Fomalont}, E.~B., {Weistrop}, D., \& {Wall}, J. 1986,
  Highlights of Astronomy, 7, 367

\bibitem[{{Kenter} {et~al.}(2005){Kenter}, {Murray}, {Forman}, {Jones},
  {Green}, {Kochanek}, {Vikhlinin}, {Fabricant}, {Fazio}, {Brand}, {Brown},
  {Dey}, {Jannuzi}, {Najita}, {McNamara}, {Shields}, \&
  {Rieke}}]{2005ApJS..161....9K}
{Kenter}, A., {Murray}, S.~S., {Forman}, W.~R., {et~al.} 2005, \apjs, 161, 9

\bibitem[{{Magliocchetti} {et~al.}(1998){Magliocchetti}, {Maddox}, {Lahav}, \&
  {Wall}}]{1998MNRAS.300..257M}
{Magliocchetti}, M., {Maddox}, S.~J., {Lahav}, O., \& {Wall}, J.~V. 1998,
  \mnras, 300, 257

\bibitem[{{Mahony} {et~al.}(2016){Mahony}, {Morganti}, {Prandoni}, {van
  Bemmel}, {Shimwell}, {Brienza}, {Best}, {Br{\"u}ggen}, {Calistro Rivera}, {de
  Gasperin}, {Hardcastle}, {Harwood}, {Heald}, {Jarvis}, {Mandal}, {Miley},
  {Retana-Montenegro}, {R{\"o}ttgering}, {Sabater}, {Tasse}, {van Velzen}, {van
  Weeren}, {Williams}, \& {White}}]{2016MNRAS.463.2997M}
{Mahony}, E.~K., {Morganti}, R., {Prandoni}, I., {et~al.} 2016, \mnras, 463,
  2997

\bibitem[{{Mandal}(in prep.)}]{Mandal2018}
{Mandal}, S. in prep.

\bibitem[{{Miller} {et~al.}(2013){Miller}, {Bonzini}, {Fomalont}, {Kellermann},
  {Mainieri}, {Padovani}, {Rosati}, {Tozzi}, \&
  {Vattakunnel}}]{2013ApJS..205...13M}
{Miller}, N.~A., {Bonzini}, M., {Fomalont}, E.~B., {et~al.} 2013, \apjs, 205,
  13

\bibitem[{{Mohan} \& {Rafferty}(2015)}]{2015ascl.soft02007M}
{Mohan}, N. \& {Rafferty}, D. 2015, {PyBDSM: Python Blob Detection and Source
  Measurement}, Astrophysics Source Code Library

\bibitem[{{Noordam}(2004)}]{2004SPIE.5489..817N}
{Noordam}, J.~E. 2004, in \procspie, Vol. 5489, Ground-based Telescopes, ed.
  J.~M. {Oschmann}, Jr., 817--825

\bibitem[{{Offringa} {et~al.}(2010){Offringa}, {de Bruyn}, {Biehl}, {Zaroubi},
  {Bernardi}, \& {Pandey}}]{2010MNRAS.405..155O}
{Offringa}, A.~R., {de Bruyn}, A.~G., {Biehl}, M., {et~al.} 2010, \mnras, 405,
  155

\bibitem[{{Offringa} {et~al.}(2014){Offringa}, {McKinley}, {Hurley-Walker},
  {Briggs}, {Wayth}, {Kaplan}, {Bell}, {Feng}, {Neben}, {Hughes}, {Rhee},
  {Murphy}, {Bhat}, {Bernardi}, {Bowman}, {Cappallo}, {Corey}, {Deshpande},
  {Emrich}, {Ewall-Wice}, {Gaensler}, {Goeke}, {Greenhill}, {Hazelton},
  {Hindson}, {Johnston-Hollitt}, {Jacobs}, {Kasper}, {Kratzenberg}, {Lenc},
  {Lonsdale}, {Lynch}, {McWhirter}, {Mitchell}, {Morales}, {Morgan},
  {Kudryavtseva}, {Oberoi}, {Ord}, {Pindor}, {Procopio}, {Prabu}, {Riding},
  {Roshi}, {Shankar}, {Srivani}, {Subrahmanyan}, {Tingay}, {Waterson},
  {Webster}, {Whitney}, {Williams}, \& {Williams}}]{2014MNRAS.444..606O}
{Offringa}, A.~R., {McKinley}, B., {Hurley-Walker}, N., {et~al.} 2014, \mnras,
  444, 606

\bibitem[{{Offringa} {et~al.}(2012){Offringa}, {van de Gronde}, \&
  {Roerdink}}]{2012A&A...539A..95O}
{Offringa}, A.~R., {van de Gronde}, J.~J., \& {Roerdink}, J.~B.~T.~M. 2012,
  \aap, 539, A95

\bibitem[{{Padovani}(2011)}]{2011MNRAS.411.1547P}
{Padovani}, P. 2011, \mnras, 411, 1547

\bibitem[{{Padovani} {et~al.}(2015){Padovani}, {Bonzini}, {Kellermann},
  {Miller}, {Mainieri}, \& {Tozzi}}]{2015MNRAS.452.1263P}
{Padovani}, P., {Bonzini}, M., {Kellermann}, K.~I., {et~al.} 2015, \mnras, 452,
  1263

\bibitem[{{Padovani} {et~al.}(2009){Padovani}, {Mainieri}, {Tozzi},
  {Kellermann}, {Fomalont}, {Miller}, {Rosati}, \&
  {Shaver}}]{2009ApJ...694..235P}
{Padovani}, P., {Mainieri}, V., {Tozzi}, P., {et~al.} 2009, \apj, 694, 235

\bibitem[{{Padovani} {et~al.}(2011){Padovani}, {Miller}, {Kellermann},
  {Mainieri}, {Rosati}, \& {Tozzi}}]{2011ApJ...740...20P}
{Padovani}, P., {Miller}, N., {Kellermann}, K.~I., {et~al.} 2011, \apj, 740, 20

\bibitem[{{Prandoni} {et~al.}(2000){Prandoni}, {Gregorini}, {Parma}, {de
  Ruiter}, {Vettolani}, {Wieringa}, \& {Ekers}}]{2000A&AS..146...41P}
{Prandoni}, I., {Gregorini}, L., {Parma}, P., {et~al.} 2000, \aaps, 146, 41

\bibitem[{{Prandoni} {et~al.}(2001){Prandoni}, {Gregorini}, {Parma}, {de
  Ruiter}, {Vettolani}, {Wieringa}, \& {Ekers}}]{2001A&A...365..392P}
{Prandoni}, I., {Gregorini}, L., {Parma}, P., {et~al.} 2001, \aap, 365, 392

\bibitem[{{Rengelink} {et~al.}(1997){Rengelink}, {Tang}, {de Bruyn}, {Miley},
  {Bremer}, {Roettgering}, \& {Bremer}}]{1997A&AS..124..259R}
{Rengelink}, R.~B., {Tang}, Y., {de Bruyn}, A.~G., {et~al.} 1997, \aaps, 124

\bibitem[{Retana-Montenegro \& R\"{o}ttgering(2018)}]{10.3389/fspas.2018.00005}
Retana-Montenegro, E. \& R\"{o}ttgering, H. 2018, Frontiers in Astronomy and
  Space Sciences, 5, 5

\bibitem[{{R{\"o}ttgering} {et~al.}(2011){R{\"o}ttgering}, {Afonso}, {Barthel},
  {Batejat}, {Best}, {Bonafede}, {Br{\"u}ggen}, {Brunetti}, {Chy{\.z}y},
  {Conway}, {Gasperin}, {Ferrari}, {Haverkorn}, {Heald}, {Hoeft}, {Jackson},
  {Jarvis}, {Ker}, {Lehnert}, {Macario}, {McKean}, {Miley}, {Morganti},
  {Oosterloo}, {Orr{\`u}}, {Pizzo}, {Rafferty}, {Shulevski}, {Tasse}, {Bemmel},
  {van der Tol}, {van Weeren}, {Verheijen}, {White}, \&
  {Wise}}]{2011JApA...32..557R}
{R{\"o}ttgering}, H., {Afonso}, J., {Barthel}, P., {et~al.} 2011, Journal of
  Astrophysics and Astronomy, 32, 557

\bibitem[{{Sabater}(in prep.)}]{Sabater2018}
{Sabater}, J. in prep.

\bibitem[{{Scaife} \& {Heald}(2012)}]{2012MNRAS.423L..30S}
{Scaife}, A.~M.~M. \& {Heald}, G.~H. 2012, \mnras, 423, L30

\bibitem[{{Schinnerer} {et~al.}(2010){Schinnerer}, {Sargent}, {Bondi}, {Smol{\v
  c}i{\'c}}, {Datta}, {Carilli}, {Bertoldi}, {Blain}, {Ciliegi}, {Koekemoer},
  \& {Scoville}}]{2010ApJS..188..384S}
{Schinnerer}, E., {Sargent}, M.~T., {Bondi}, M., {et~al.} 2010, \apjs, 188, 384

\bibitem[{{Schwab}(1984)}]{1984AJ.....89.1076S}
{Schwab}, F.~R. 1984, \aj, 89, 1076

\bibitem[{{Shimwell} {et~al.}(2017){Shimwell}, {R{\"o}ttgering}, {Best},
  {Williams}, {Dijkema}, {de Gasperin}, {Hardcastle}, {Heald}, {Hoang},
  {Horneffer}, {Intema}, {Mahony}, {Mandal}, {Mechev}, {Morabito}, {Oonk},
  {Rafferty}, {Retana-Montenegro}, {Sabater}, {Tasse}, {van Weeren},
  {Br{\"u}ggen}, {Brunetti}, {Chy{\.z}y}, {Conway}, {Haverkorn}, {Jackson},
  {Jarvis}, {McKean}, {Miley}, {Morganti}, {White}, {Wise}, {van Bemmel},
  {Beck}, {Brienza}, {Bonafede}, {Calistro Rivera}, {Cassano}, {Clarke},
  {Cseh}, {Deller}, {Drabent}, {van Driel}, {Engels}, {Falcke}, {Ferrari},
  {Fr{\"o}hlich}, {Garrett}, {Harwood}, {Heesen}, {Hoeft}, {Horellou},
  {Israel}, {Kapi{\'n}ska}, {Kunert-Bajraszewska}, {McKay}, {Mohan},
  {Orr{\'u}}, {Pizzo}, {Prandoni}, {Schwarz}, {Shulevski}, {Sipior}, {Smith},
  {Sridhar}, {Steinmetz}, {Stroe}, {Varenius}, {van der Werf}, {Zensus}, \&
  {Zwart}}]{2017AA...598A.104S}
{Shimwell}, T.~W., {R{\"o}ttgering}, H.~J.~A., {Best}, P.~N., {et~al.} 2017,
  \aap, 598, A104

\bibitem[{{Smirnov}(2011)}]{2011A&A...527A.107S}
{Smirnov}, O.~M. 2011, \aap, 527, A107

\bibitem[{{Smirnov} \& {Tasse}(2015)}]{2015MNRAS.449.2668S}
{Smirnov}, O.~M. \& {Tasse}, C. 2015, \mnras, 449, 2668

\bibitem[{{Smol{\v c}i{\'c}} {et~al.}(2017{\natexlab{a}}){Smol{\v c}i{\'c}},
  {Delvecchio}, {Zamorani}, {Baran}, {Novak}, {Delhaize}, {Schinnerer},
  {Berta}, {Bondi}, {Ciliegi}, {Capak}, {Civano}, {Karim}, {Le Fevre},
  {Ilbert}, {Laigle}, {Marchesi}, {McCracken}, {Tasca}, {Salvato}, \&
  {Vardoulaki}}]{2017A&A...602A...2S}
{Smol{\v c}i{\'c}}, V., {Delvecchio}, I., {Zamorani}, G., {et~al.}
  2017{\natexlab{a}}, \aap, 602, A2

\bibitem[{{Smol{\v c}i{\'c}} {et~al.}(2017{\natexlab{b}}){Smol{\v c}i{\'c}},
  {Novak}, {Bondi}, {Ciliegi}, {Mooley}, {Schinnerer}, {Zamorani}, {Navarrete},
  {Bourke}, {Karim}, {Vardoulaki}, {Leslie}, {Delhaize}, {Carilli}, {Myers},
  {Baran}, {Delvecchio}, {Miettinen}, {Banfield}, {Balokovi{\'c}}, {Bertoldi},
  {Capak}, {Frail}, {Hallinan}, {Hao}, {Herrera Ruiz}, {Horesh}, {Ilbert},
  {Intema}, {Jeli{\'c}}, {Kl{\"o}ckner}, {Krpan}, {Kulkarni}, {McCracken},
  {Laigle}, {Middleberg}, {Murphy}, {Sargent}, {Scoville}, \&
  {Sheth}}]{2017A&A...602A...1S}
{Smol{\v c}i{\'c}}, V., {Novak}, M., {Bondi}, M., {et~al.} 2017{\natexlab{b}},
  \aap, 602, A1

\bibitem[{{Smol{\v c}i{\'c}} {et~al.}(2008){Smol{\v c}i{\'c}}, {Schinnerer},
  {Scodeggio}, {Franzetti}, {Aussel}, {Bondi}, {Brusa}, {Carilli}, {Capak},
  {Charlot}, {Ciliegi}, {Ilbert}, {Ivezi{\'c}}, {Jahnke}, {McCracken},
  {Obri{\'c}}, {Salvato}, {Sanders}, {Scoville}, {Trump}, {Tremonti}, {Tasca},
  {Walcher}, \& {Zamorani}}]{2008ApJS..177...14S}
{Smol{\v c}i{\'c}}, V., {Schinnerer}, E., {Scodeggio}, M., {et~al.} 2008,
  \apjs, 177, 14

\bibitem[{{Tasse}(2014)}]{2014arXiv1410.8706T}
{Tasse}, C. 2014, ArXiv e-prints, arXiv:1410.8706

\bibitem[{{Tasse}(in prep.)}]{Tasse2018}
{Tasse}, C. in prep.

\bibitem[{{Tasse} {et~al.}(2017){Tasse}, {Hugo}, {Mirmont}, {Smirnov},
  {Atemkeng}, {Bester}, {Hardcastle}, {Lakhoo}, {Perkins}, \&
  {Shimwell}}]{2017arXiv171202078T}
{Tasse}, C., {Hugo}, B., {Mirmont}, M., {et~al.} 2017, ArXiv e-prints,
  arXiv:1712.02078

\bibitem[{{van Weeren} {et~al.}(2016){van Weeren}, {Williams}, {Hardcastle},
  {Shimwell}, {Rafferty}, {Sabater}, {Heald}, {Sridhar}, {Dijkema}, {Brunetti},
  {Br{\"u}ggen}, {Andrade-Santos}, {Ogrean}, {R{\"o}ttgering}, {Dawson},
  {Forman}, {de Gasperin}, {Jones}, {Miley}, {Rudnick}, {Sarazin}, {Bonafede},
  {Best}, {B{\^i}rzan}, {Cassano}, {Chy{\.z}y}, {Croston}, {Ensslin},
  {Ferrari}, {Hoeft}, {Horellou}, {Jarvis}, {Kraft}, {Mevius}, {Intema},
  {Murray}, {Orr{\'u}}, {Pizzo}, {Simionescu}, {Stroe}, {van der Tol}, \&
  {White}}]{2016ApJS..223....2V}
{van Weeren}, R.~J., {Williams}, W.~L., {Hardcastle}, M.~J., {et~al.} 2016,
  \apjs, 223, 2

\bibitem[{{van Weeren} {et~al.}(2014){van Weeren}, {Williams}, {Tasse},
  {R{\"o}ttgering}, {Rafferty}, {van der Tol}, {Heald}, {White}, {Shulevski},
  {Best}, {Intema}, {Bhatnagar}, {Reich}, {Steinmetz}, {van Velzen},
  {En{\ss}lin}, {Prandoni}, {de Gasperin}, {Jamrozy}, {Brunetti}, {Jarvis},
  {McKean}, {Wise}, {Ferrari}, {Harwood}, {Oonk}, {Hoeft},
  {Kunert-Bajraszewska}, {Horellou}, {Wucknitz}, {Bonafede}, {Mohan}, {Scaife},
  {Kl{\"o}ckner}, {van Bemmel}, {Merloni}, {Chyzy}, {Engels}, {Falcke},
  {Pandey-Pommier}, {Alexov}, {Anderson}, {Avruch}, {Beck}, {Bell}, {Bentum},
  {Bernardi}, {Breitling}, {Broderick}, {Brouw}, {Br{\"u}ggen}, {Butcher},
  {Ciardi}, {de Geus}, {de Vos}, {Deller}, {Duscha}, {Eisl{\"o}ffel},
  {Fallows}, {Frieswijk}, {Garrett}, {Grie{\ss}meier}, {Gunst}, {Hamaker},
  {Hassall}, {H{\"o}randel}, {van der Horst}, {Iacobelli}, {Jackson}, {Juette},
  {Kondratiev}, {Kuniyoshi}, {Maat}, {Mann}, {McKay-Bukowski}, {Mevius},
  {Morganti}, {Munk}, {Offringa}, {Orr{\`u}}, {Paas}, {Pandey}, {Pietka},
  {Pizzo}, {Polatidis}, {Renting}, {Rowlinson}, {Schwarz}, {Serylak}, {Sluman},
  {Smirnov}, {Stappers}, {Stewart}, {Swinbank}, {Tagger}, {Tang}, {Thoudam},
  {Toribio}, {Vermeulen}, {Vocks}, \& {Zarka}}]{2014ApJ...793...82V}
{van Weeren}, R.~J., {Williams}, W.~L., {Tasse}, C., {et~al.} 2014, \apj, 793,
  82

\bibitem[{{Vernstrom} {et~al.}(2016){Vernstrom}, {Scott}, {Wall}, {Condon},
  {Cotton}, {Kellermann}, \& {Perley}}]{2016MNRAS.462.2934V}
{Vernstrom}, T., {Scott}, D., {Wall}, J.~V., {et~al.} 2016, \mnras, 462, 2934

\bibitem[{{Wayth} {et~al.}(2015){Wayth}, {Lenc}, {Bell}, {Callingham},
  {Dwarakanath}, {Franzen}, {For}, {Gaensler}, {Hancock}, {Hindson},
  {Hurley-Walker}, {Jackson}, {Johnston-Hollitt}, {Kapi{\'n}ska}, {McKinley},
  {Morgan}, {Offringa}, {Procopio}, {Staveley-Smith}, {Wu}, {Zheng}, {Trott},
  {Bernardi}, {Bowman}, {Briggs}, {Cappallo}, {Corey}, {Deshpande}, {Emrich},
  {Goeke}, {Greenhill}, {Hazelton}, {Kaplan}, {Kasper}, {Kratzenberg},
  {Lonsdale}, {Lynch}, {McWhirter}, {Mitchell}, {Morales}, {Morgan}, {Oberoi},
  {Ord}, {Prabu}, {Rogers}, {Roshi}, {Shankar}, {Srivani}, {Subrahmanyan},
  {Tingay}, {Waterson}, {Webster}, {Whitney}, {Williams}, \&
  {Williams}}]{2015PASA...32...25W}
{Wayth}, R.~B., {Lenc}, E., {Bell}, M.~E., {et~al.} 2015, \pasa, 32, e025

\bibitem[{{White} {et~al.}(2012){White}, {Hatsukade}, {Pearson}, {Takagi},
  {Sedgwick}, {Matsuura}, {Matsuhara}, {Serjeant}, {Nakagawa}, {Lee}, {Oyabu},
  {Jeong}, {Shirahata}, {Kohno}, {Yamamura}, {Hanami}, {Goto}, {Makiuti},
  {Clements}, {Malek}, \& {Khan}}]{2012MNRAS.427.1830W}
{White}, G.~J., {Hatsukade}, B., {Pearson}, C., {et~al.} 2012, \mnras, 427,
  1830

\bibitem[{{Williams} {et~al.}(2013){Williams}, {Intema}, \&
  {R{\"o}ttgering}}]{2013AA...549A..55W}
{Williams}, W.~L., {Intema}, H.~T., \& {R{\"o}ttgering}, H.~J.~A. 2013, \aap,
  549, A55

\bibitem[{{Williams} {et~al.}(2016){Williams}, {van Weeren}, {R{\"o}ttgering},
  {Best}, {Dijkema}, {de Gasperin}, {Hardcastle}, {Heald}, {Prandoni},
  {Sabater}, {Shimwell}, {Tasse}, {van Bemmel}, {Br{\"u}ggen}, {Brunetti},
  {Conway}, {En{\ss}lin}, {Engels}, {Falcke}, {Ferrari}, {Haverkorn},
  {Jackson}, {Jarvis}, {Kapi{\'n}ska}, {Mahony}, {Miley}, {Morabito},
  {Morganti}, {Orr{\'u}}, {Retana-Montenegro}, {Sridhar}, {Toribio}, {White},
  {Wise}, \& {Zwart}}]{2016MNRAS.460.2385W}
{Williams}, W.~L., {van Weeren}, R.~J., {R{\"o}ttgering}, H.~J.~A., {et~al.}
  2016, \mnras, 460, 2385

\bibitem[{{Wilman} {et~al.}(2008){Wilman}, {Miller}, {Jarvis}, {Mauch},
  {Levrier}, {Abdalla}, {Rawlings}, {Kl{\"o}ckner}, {Obreschkow}, {Olteanu}, \&
  {Young}}]{2008MNRAS.388.1335W}
{Wilman}, R.~J., {Miller}, L., {Jarvis}, M.~J., {et~al.} 2008, \mnras, 388,
  1335

\bibitem[{{Windhorst} {et~al.}(1990){Windhorst}, {Mathis}, \&
  {Neuschaefer}}]{1990ASPC...10..389W}
{Windhorst}, R., {Mathis}, D., \& {Neuschaefer}, L. 1990, in Astronomical
  Society of the Pacific Conference Series, Vol.~10, Evolution of the Universe
  of Galaxies, ed. R.~G. {Kron}, 389--403

\bibitem[{{Windhorst} {et~al.}(1993){Windhorst}, {Fomalont}, {Partridge}, \&
  {Lowenthal}}]{1993ApJ...405..498W}
{Windhorst}, R.~A., {Fomalont}, E.~B., {Partridge}, R.~B., \& {Lowenthal},
  J.~D. 1993, \apj, 405, 498

\bibitem[{{Windhorst} {et~al.}(1985){Windhorst}, {Miley}, {Owen}, {Kron}, \&
  {Koo}}]{1985ApJ...289..494W}
{Windhorst}, R.~A., {Miley}, G.~K., {Owen}, F.~N., {Kron}, R.~G., \& {Koo},
  D.~C. 1985, \apj, 289, 494

\bibitem[{{Yatawatta} {et~al.}(2013){Yatawatta}, {de Bruyn}, {Brentjens},
  {Labropoulos}, {Pandey}, {Kazemi}, {Zaroubi}, {Koopmans}, {Offringa},
  {Jeli{\'c}}, {Martinez Rubi}, {Veligatla}, {Wijnholds}, {Brouw}, {Bernardi},
  {Ciardi}, {Daiboo}, {Harker}, {Mellema}, {Schaye}, {Thomas}, {Vedantham},
  {Chapman}, {Abdalla}, {Alexov}, {Anderson}, {Avruch}, {Batejat}, {Bell},
  {Bell}, {Bentum}, {Best}, {Bonafede}, {Bregman}, {Breitling}, {van de Brink},
  {Broderick}, {Br{\"u}ggen}, {Conway}, {de Gasperin}, {de Geus}, {Duscha},
  {Falcke}, {Fallows}, {Ferrari}, {Frieswijk}, {Garrett}, {Griessmeier},
  {Gunst}, {Hassall}, {Hessels}, {Hoeft}, {Iacobelli}, {Juette},
  {Karastergiou}, {Kondratiev}, {Kramer}, {Kuniyoshi}, {Kuper}, {van Leeuwen},
  {Maat}, {Mann}, {McKean}, {Mevius}, {Mol}, {Munk}, {Nijboer}, {Noordam},
  {Norden}, {Orru}, {Paas}, {Pandey-Pommier}, {Pizzo}, {Polatidis}, {Reich},
  {R{\"o}ttgering}, {Sluman}, {Smirnov}, {Stappers}, {Steinmetz}, {Tagger},
  {Tang}, {Tasse}, {ter Veen}, {Vermeulen}, {van Weeren}, {Wise}, {Wucknitz},
  \& {Zarka}}]{2013A&A...550A.136Y}
{Yatawatta}, S., {de Bruyn}, A.~G., {Brentjens}, M.~A., {et~al.} 2013, \aap,
  550, A136

\end{thebibliography}
.
\end{document}